\documentclass[11pt,aps,tightenlines,nofootinbib]{revtex4}
\usepackage{amsmath,amscd,amssymb}
\usepackage[usenames,dvipsnames]{color}
\usepackage{epsfig}
\usepackage{xfrac}
\usepackage{ulem}
\usepackage{latexsym}
\usepackage{mathrsfs}
\usepackage{graphicx}
\usepackage{bbm}      
\def\beq{\begin{equation}}
\def\eeq{\end{equation}}
\def\be{\begin{eqnarray}}
\def\ee{\end{eqnarray}}
\newcommand{\dslash}{\partial \hskip -0.6em /}
\newcommand{\Dslash}{D \hskip -0.6em /}

\newcommand{\zr}[1]{\mbox{\hspace*{#1em}}}

\newcommand{\tr}{\mbox{tr}}

\newcommand{\Vek}[1]{\mbox{\boldmath$#1$\unboldmath}}
\newcommand{\vek}[1]{\mathbf{#1}}

\newcommand{\ID}{\mbox{{\sf 1}\zr{-0.16}\rule{0.04em}{1.55ex}\zr{0.1}}}
\newcommand{\ar}{\alpha_r}
\newcommand{\ag}{\alpha_G}
\newcommand{\ax}{\alpha_\xi}
\newcommand{\ah}{\alpha_H}
\newcommand{\alp}{\alpha_P}
\newcommand{\sx}{s_\xi}
\newcommand{\sd}{s_\Delta}
\newcommand{\cx}{c_\xi}
\newcommand{\cd}{c_\Delta}
\newcommand{\sy}{s_{\xi_1}}
\newcommand{\cy}{c_{\xi_1}}

\newcommand{\phiunit}{\widehat{\Vek{\varphi}}}
\newcommand{\rhounit}{\widehat{\Vek{\rho}}}

\begin{document}

\title{Fermion Energies in the Background of a Cosmic String}

\author{N. Graham$^{a)}$, M. Quandt$^{b)}$, H. Weigel$^{c)}$}

\affiliation{
$^{a)}$Department of Physics, Middlebury College
Middlebury, VT 05753, USA\\
$^{b)}$Institute for Theoretical Physics, T\"ubingen University
D--72076 T\"ubingen, Germany\\
$^{c)}$Physics Department, Stellenbosch University,
Matieland 7602, South Africa}

\begin{abstract}
We provide a thorough exposition, including technical and numerical
details, of previously published results on the quantum stabilization
of cosmic strings. Stabilization occurs through the coupling to a heavy
fermion doublet in a reduced version of the standard model. We
combine the vacuum polarization energy
of fermion zero--point fluctuations and the binding energy of occupied
energy levels, which are of the same order in a semi--classical
expansion. Populating these bound states assigns a charge to the
string.  We show that strings carrying
fermion charge become stable if the electro--weak bosons are coupled to a
fermion that is less than twice as heavy as the top quark.
The vacuum remains stable in our model, because neutral strings are
not energetically favored. These findings suggests that extraordinarily
large fermion masses or unrealistic couplings are not required to bind a
cosmic string in the standard model.
\end{abstract}

\maketitle

\section{Introduction}
It is well--known that the electroweak standard model and many of its
extensions have the potential to support string--like
configurations that are the particle physics analogs
of vortices or  magnetic flux tubes in condensed matter physics. Such
objects are usually  called \emph{cosmic strings} to distinguish them
from the fundamental variables in string theory, and also to indicate
that they typically stretch over cosmic length scales.

The topology of string--like configurations is described by
the first homotopy group $\Pi_1(\mathscr{M})$, where $\mathscr{M}$ is the
manifold of vacuum field configurations far away from the string.
In typical electroweak--like models, a Higgs condensate breaks an
initial gauge group $G$ down to some subgroup $H$, so that
$\mathscr{M} \simeq G/H$. Topologically stable strings are therefore ruled out
in the electroweak standard model $SU(2) \times U(1) \to U(1)$ because
$G/H$ is simply connected. Nevertheless, one could
envision a GUT  and/or supersymmetric extension in which a simply
connected group $G$ breaks down to the electroweak $SU(2) \times U(1)$
at a much higher scale, so that
$\Pi_1(G / (SU(2)\times U(1)))$ is nontrivial. Since such GUT strings would
have enormous energy densities, they could be seen by direct observation using
gravitational lensing \cite{Kibble:1976sj,Hindmarsh:1994re}
or by signatures in the cosmic microwave background \cite{Vilenkin:1994}.
Moreover, a network of  such strings is a candidate for the
dark energy required to explain the recently observed cosmic
acceleration \cite{Perlmutter:1998np,Riess:1998cb}.

The absence of topological stability does not imply that electroweak strings
(or $Z$--\emph{strings} \cite{Vachaspati:1992fi,Achucarro:1999it,Nambu:1977ag})
are unstable or irrelevant for particle physics. While their direct gravitational 
effects are negligible, $Z$--strings can still be relevant for cosmology 
at a sub--dominant level \cite{Copeland:2009ga,Achucarro:2008fn}.
Their most interesting consequences originate, however, from their
coupling to the standard model fields. $Z$--strings provide a source
for primordial magnetic fields \cite{Nambu:1977ag}
and they also offer a scenario for baryogenesis with a second order
phase transition \cite{Brandenberger:1992ys,Brandenberger:1994bx}.
In contrast, a strong first order transition as required by the usual bubble
nucleation scenario is unlikely in the electroweak standard model \cite{EWPhase},
without non-standard additions such as supersymmetry or higher--dimensional
operators \cite{Grojean:2004xa}.
Because the core of the $Z$--string is characterized by a suppressed Higgs condensate,
it allows for both the copious baryon number violation and the
out--of equilibrium regions required by the Sakharov conditions, without
relying on a first order phase transition.

However, these interesting effects are only viable if $Z$--strings are
energetically stabilized by their coupling to the remaining quantum fields.
The most important contributions are expected to come from (heavy) fermions,
since their quantum energy dominates in the limit $N_C \to \infty$, where
$N_C$ is the number of QCD colors or other internal degrees of freedom.
The Dirac spectrum in typical string backgrounds is deformed to contain either
an exact or near zero mode, so that fermions can substantially lower their
energy by binding to the string. This binding effect can overcome the
classical energy required to form the string background. However, the
remaining spectrum of modes is also deformed and for consistency
its contribution (the vacuum polarization energy) must be taken into account as
well. Heavier fermions are expected to provide more binding since the
energy gain per fermion charge is higher; a similar conclusion can
also be obtained from decoupling arguments
\cite{D'Hoker:1984ph,D'Hoker:1984pc}. Dynamical stability of
$Z$--strings in the full standard model also would suggest that
they are presently observable.

A number of previous studies have investigated quantum properties of string
configurations. Naculich~\cite{Naculich:1995cb} has shown that in the
limit of weak coupling, fermion fluctuations destabilize the string.
The quantum properties of $Z$--strings have also been connected to
non--perturbative anomalies~\cite{Klinkhamer:2003hz}.
The emergence or absence of exact neutrino zero modes in a
$Z$--string background and the possible consequences for the
string topology were investigated in \cite{Stojkovic}.
A first attempt at a full calculation of the fermionic quantum corrections
to the $Z$--string energy was carried out in ref.~\cite{Groves:1999ks}.
Those authors were only able to compare the energies of two
string configurations, rather than comparing a single string
configuration to the vacuum because of limitations arising from the non--trivial
behavior at spatial infinity which we discuss below.
The fermionic vacuum polarization energy of the Abelian Nielsen--Olesen 
vortex~\cite{Nielsen:1973cs} has been estimated in ref.~\cite{Bordag:2003at}
with regularization limited to the subtraction of the divergences
in the heat--kernel expansion. Quantum energies of bosonic fluctuations
in string backgrounds were calculated in ref.~\cite{Baacke:2008sq}.
Finally, the dynamical fields coupled to the string can also result in
(Abelian or non--Abelian) currents running along the string's core.
The time evolution of such structured strings was studied in
ref.~\cite{Lilley:2010av}, where the current was induced by the
coupling to an extra scalar field.

We have previously pursued the idea of stabilizing
cosmic strings by  populating fermionic bound states in a $2+1$ dimensional
model \cite{Graham:2006qt}. Many such bound states emerge including, in some configurations,
an exact zero--mode~\cite{Naculich:1995cb}.
Nonetheless, stable configurations were only obtained for extreme values of
the model parameters. In $3+1$ dimensions, stability is potentially easier to achieve
because quantization of the momentum parallel to the symmetry axis
supplies an additional multiplicity of bound states.

In this paper, we will employ the phase--shift approach, or spectral
method, to compute the complete $\mathcal{O}(\hbar)$ fermionic 
contribution to the string energy from first principles.  This is not a 
simple task, since the string has a vortex structure that introduces 
non--trivial field winding at spatial infinity. The  standard spectral 
methods are thus not directly applicable since scattering theory off the 
string is ill--defined. More precisely, the Born expansion to the 
vacuum polarization energy, which in the phase shift approach is 
identified with the Feynman series, does not exist for the string 
background in its standard formulation. Recently we have 
shown how to overcome these problems by choosing a particular set of
gauges~\cite{Weigel:2009wi,Weigel:2010pf}. Numerical results of the full
calculation of the string's quantum energy were first reported in
ref.~\cite{Weigel:2010zk}.  Here we will present the technical details
of our calculation along with improved numerical data and a discussion
of possible consequences of our finding.

This paper is organized as follows: In the next section we describe our 
model and the string configuration. We then discuss the fermion Hamiltonian 
of our model and, in particular, how a local gauge transformation can be 
used to solve the technical problem of long--ranged gauge potentials in the 
string background. We also present the grand spin decomposition of the 
scattering problem.
Section \ref{sec:4} gives a detailed account of our
method for computing the fermion quantum energy, which is based on the spectral
approach \cite{Graham:2009zz} and the interface formalism \cite{Graham:2001dy}.
The individual contributions to the string's quantum
energy are described in separate subsections, while some (lengthy) numerical
details are deferred to appendices \ref{appA}, \ref{appB}, and \ref{appC}.
The omission of boson fluctuations causes the model not to be 
asymptotically free which then introduces an unphysical Landau pole. In 
appendix D we verify that our results for the vacuum polarization energy
are not affected by this artifact.
In section \ref{sec:5}, we explain our variational search for a stable
string configuration. To occupy fermion levels in the string
background, we introduce a quantity similar to the chemical potential
in statistical mechanics, which allows us to compute
the total binding energy of the string as a function of the prescribed
fermion charge of the string.

Our numerical results are presented in detail within section \ref{sec:6}.
We show the parameter dependence of the individual contributions to
the string's quantum energy. The stable configuration is discussed
in more detail and it is shown that for the most stable configuration
the gauge field contribution is negligible compared to the
deformation of the Higgs field. Stabilization
occurs for otherwise realistic parameters if the Yukawa coupling is increased
by about 70\% from the value for the top quark. Since we keep all other
parameters as suggested by the standard model, this corresponds to a
fermion mass of about $300\,\mathrm{GeV}$.
We close in section \ref{sec:7} with a brief summary of our results,
a discussion of its implication for the electroweak standard model and
an outlook on possible directions for future work.

We have published some of the results earlier~\cite{Weigel:2010zk} and
therefore focus on the technical aspects of the calculation here.

\section{The model}
\label{sec:2}

We consider a left--handed $SU(2)$ gauge theory in which a fermion
doublet
$\Psi=\begin{pmatrix} \Psi_t \cr \Psi_b\end{pmatrix}$
is coupled to a triplet gauge field
$W_\mu=\frac{1}{2}\begin{pmatrix}W^0_\mu &\sqrt{2}W^+_\mu\\[2mm]
\sqrt{2}W^-_\mu & -W^0_\mu\end{pmatrix}$ and a Higgs doublet
$\phi = \begin{pmatrix} \phi_+ \cr \phi_0 \end{pmatrix}$.
Both components, $\Psi_t$ and $\Psi_b$, are Dirac four--spinors.
This model is intended to represent the electroweak interactions, where we
introduce some technical modifications to simplify the analysis:

\begin{enumerate}
\item we set the Weinberg angle to zero so that electromagnetism decouples and
the gauge bosons become degenerate in mass;
\item we neglect QCD interactions, although the color degeneracy, $N_C =3$, is
included in the quantum energy arising from the fermions;
\item we only consider a single fermion doublet and neglect inter--species
(CKM) mixing and mass splitting within the doublet.
\end{enumerate}

\noindent
With these adjustments, the bosonic part of our model is described
by the Lagrangian
\begin{equation}
\mathcal{L}_{\phi,W}=-\frac{1}{2} \tr
\left(G^{\mu\nu}G_{\mu\nu}\right) +
\frac{1}{2} \tr \left(D^{\mu}\Phi \right)^{\dag} D_{\mu}\Phi
- \frac{\lambda}{2} \tr \left(\Phi^{\dag} \Phi - v^2 \right)^2 \,,
\label{Lbosonic}
\end{equation}
where the Higgs doublet is written using the usual matrix
representation
\[
\Phi=\begin{pmatrix}
\phi_0^* & \phi_+ \cr -\phi_+^* & \phi_0 \end{pmatrix} \,.
\]
The gauge coupling constant $g$ enters through the covariant derivative
$D_\mu = \partial_\mu - i \,g W_\mu$, and the $SU(2)$ field  strength tensor is
\begin{equation}
G_{\mu\nu} = \partial_\mu\,W_\nu - \partial_\nu W_\mu - i \,g\,[ \,W_\mu\,,\,
W_\nu\,]\,.
\label{fieldtensor}
\end{equation}
We treat the bosonic fields as a classical background, ignoring the effects
of bosonic fluctuations. This approach can be justified formally by the limit
of a large number of colors $N_C\to\infty$, even though no QCD interactions are
included: Since the quarks carry a color quantum number in the fundamental
representation of the color group $SU(N_C)$, their contribution to the quantum
energy is enhanced by a factor $N_C$ as compared to the bosonic quantum contribution.
Hence we compute the leading quantum corrections to
the classical background energy from the fermion Lagrangian
\begin{equation}
\mathcal{L}_\Psi=i\overline{\Psi}
\left(P_L \Dslash  + P_R \dslash \right) \Psi
-f\,\overline{\Psi}\left(\Phi P_R+\Phi^\dagger P_L\right)\Psi\,.
\label{gaugelag}
\end{equation}
Here, $P_{R,L}=\frac{1}{2}\left(1\pm\gamma_5\right)$ are projection
operators on left/right--handed components, respectively, and the strength
of the Higgs-fermion interaction is parameterized by the Yukawa coupling $f$,
which gives rise to the fermion mass, $m = f v$, once the Higgs acquires a
vacuum expectation value ($vev$) $v$, where $\langle {\rm det}(\Phi)\rangle = v^2 \neq 0$.
All other masses in this model are also a result of the symmetry breaking Higgs
condensate, viz.~the gauge boson mass $M_W = g v / \sqrt{2}$ and the
Higgs mass $m_H = 2 v \,\sqrt{\lambda}$.
This similarity with the standard model of
particle physics suggests the model parameters 
\begin{equation}
g=0.72\,,\quad
v=177\,{\rm GeV}\,,\quad
m_{\rm H}= 140\,{\rm GeV}\,,\quad
f=0.99\,,
\label{eq:parameters}
\end{equation}
by taking the fermion doublet to have the mass of the top quark.
Finally, the counterterm Lagrangian necessary
to renormalize the quantum energy will be listed with the computational details
in eq.~(\ref{Lcounter}) below.

As mentioned earlier, we are particularly interested in the
$Z$--string background configuration. If we consider
a single straight (infinitely extended) string along the $z$--axis, the
corresponding boson fields depend only on the planar polar
coordinates, {\it i.e.} the distance $\rho$ from the symmetry axis and the
corresponding azimuthal angle $\varphi$.  In Weyl gauge $W_0 = 0$, we have
\begin{eqnarray}
\vek{W}&=&n\,{\rm sin}(\xi_1)\,\frac{f_G(\rho)}{\rho}\,\phiunit\,
\begin{pmatrix}
{\rm sin}(\xi_1) & i \,{\rm cos}(\xi_1)\,{\rm e}^{-in\varphi} \\[2mm]
-i\, {\rm cos}(\xi_1)\,{\rm e}^{in\varphi} & - {\rm sin}(\xi_1)
\end{pmatrix}
\label{string_gauge}
\\[5mm]
\Phi&=&vf_H(\rho)
\begin{pmatrix}
{\rm sin}(\xi_1)\, {\rm e}^{-in\varphi} & -i\, {\rm cos}(\xi_1) \\[2mm]
-i \,{\rm cos}(\xi_1) & {\rm sin}(\xi_1)\, {\rm e}^{in\varphi}
\end{pmatrix}\,.
\label{string_higgs}
\end{eqnarray}
The $Z$--boson component $Z_\mu \equiv W_\mu^3$ of this configuration
has the familiar shape of an Abelian Nielsen--Olesen string of winding
number $n$, although the entire non--Abelian configuration is smoothly
deformable into the vacuum and thus not stable for any
topological reason.  We have left the analog of winding number $n$ general,
although we will only consider $n=1$ in our numerical treatment below.
The additional variational  parameter $\xi_1 \in [0,\sfrac{\pi}{2}]$ was 
introduced to include a non--trivial
gauge field in the string background; the same parameter also determines the
orientation of the Higgs field on the chiral circle. Then the
classical energy per unit length of the string is a
functional of the profile functions $f_G(\rho)$ and $f_H(\rho)$,
\begin{equation}
\frac{E_{\rm cl}}{m^2}=2\pi\int_0^\infty \rho\, d\rho\,\left\{
n^2\sin^2 \xi_1\,\biggl[\frac{2}{g^2}
\left(\frac{f_G^\prime}{\rho}\right)^2
+\frac{f_H^2}{f^2\rho^2}\,\left(1-f_G\right)^2\biggr]
+\frac{f_H^{\prime2}}{f^2}
+\frac{\mu_h^2}{4f^2}\left(1-f_H^2\right)^2\right\}\,,
\label{eq:ecl}
\end{equation}
where the radial integration variable is related to the physical radius
by $\rho_{\rm phys}=\rho/m$ and $\mu_H\equiv m_H/m$.
The radial functions $f_G(\rho)$ and $f_H(\rho)$ in the string
configuration, eqs.~(\ref{string_gauge}) and~(\ref{string_higgs})
approach unity at large distances and vanish
at the string core $\rho=0$. Typically, they will have similar shapes to the
familiar Nielsen--Olesen string, with both $\vek{W}$ and $\Phi$
going as $\mathscr{O}(\rho)$ at $\rho \to 0$ to avoid ambiguities from an
undefined azimuthal angle $\varphi$. We choose a convenient form,
\begin{equation}
f_H(\rho)=1-\exp\left[-\frac{\rho}{w_H}\right]
\quad \mbox{and} \quad
f_G(\rho)=1-\exp\left[-\left(\frac{\rho}{w_G}\right)^2\right]\,
\label{eqn:profile}
\end{equation}
with two width parameters, $w_H$ and $w_G$, which we also
measure in inverse multiples of the fermion mass $m$. Together with
the angle $\xi_1$ describing the gauge field admixture in the string,
we thus have three variational {\it ansatz} parameters, $(w_H, w_G, \xi_1)$,
in addition to the model parameters $v$ (which sets the overall scale) and
the three couplings $f,g$ and  $\lambda$, that are discussed above.

To assess the quality of the variational {\it ansatz}, eq.~(\ref{eqn:profile})
we see how well it is capable of fitting the Nielsen--Olesen profiles which 
minimize the classical energy, eq.~(\ref{eq:ecl}) for $\xi_1=\sfrac{\pi}{2}$.  
\begin{figure}[tp]
\centerline{
\includegraphics[width=10.0cm,height=5.0cm]{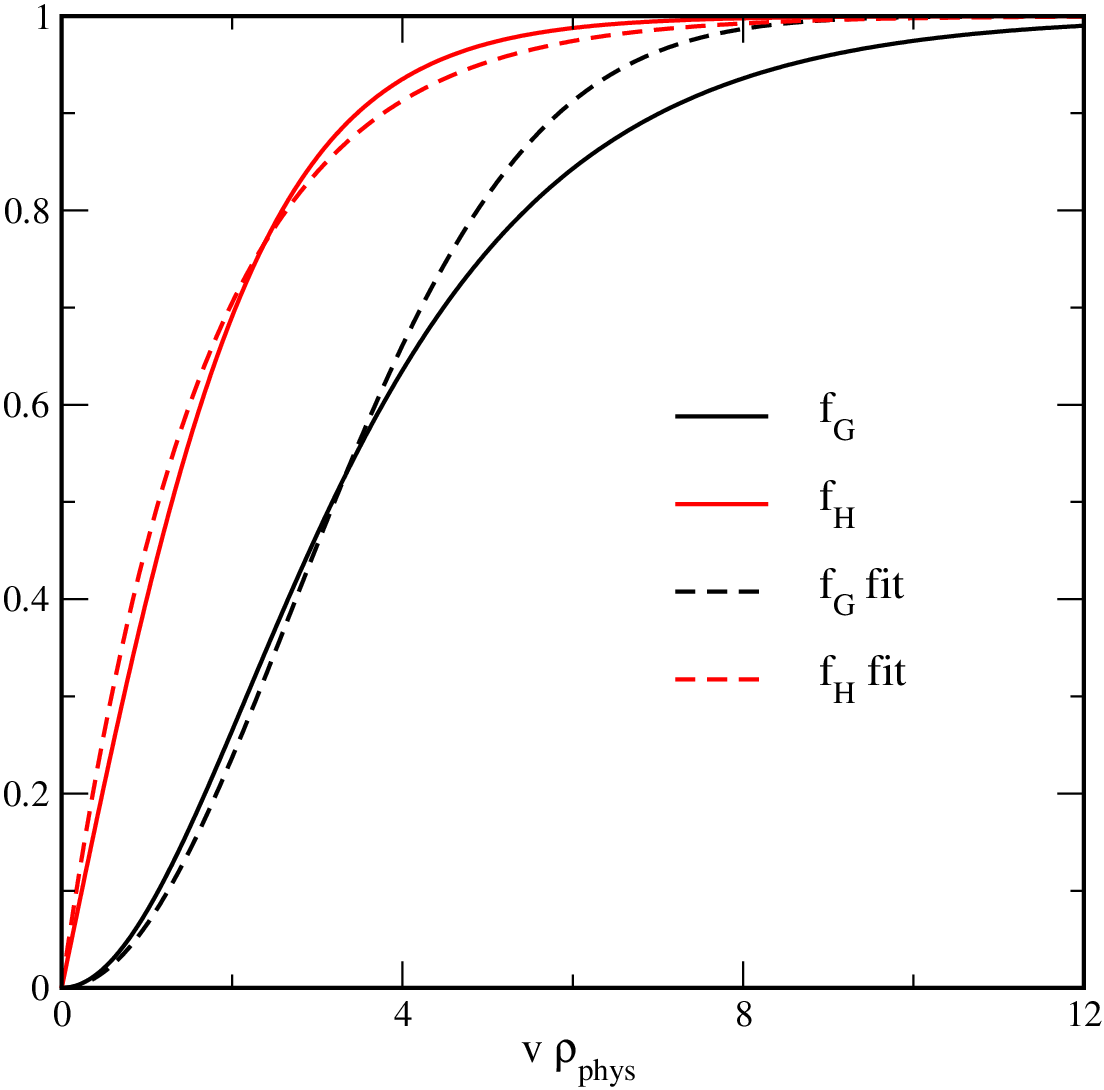}}
\caption{\label{fig:NO}(Color online) Comparison between the 
Nielsen--Olesen profiles (full lines) and a fit using the 
variational {\it ansatz} of eq.~(\ref{eqn:profile}) (dashed lines).
The variational parameters are $w_H=1.64 f$ and $w_G=3.85 f$.} 
\end{figure}
As seen from figure \ref{fig:NO} there is a minor
discrepancy at large distances for the gauge field 
profile $f_G$ due to the Gau{\ss}ian decaying faster than 
any exponential function. This discrepancy affects the result for the 
classical energy in a negligible manner. For fixed $\xi_1=\sfrac{\pi}{2}$ the 
true minimum is at $7.56v^2$ while the variational profiles yield $7.72v^2$.

\section{Dirac Hamiltonian}
The fermionic quantum corrections to the string background are computed
in several steps. First, we extract the Dirac Hamiltonian associated with the
Lagrangian eq.~(\ref{gaugelag}) and observe that the {\it ans\"atze},
eqs.~(\ref{string_gauge}) and~(\ref{string_higgs}), do not depend
on the $z$--coordinate (along the string symmetry axis). Hence this coordinate
does not appear explicitly in the Hamiltonian and the
$z$--dependence of the corresponding wave functions is simply $e^{-i p_z z}$.
To compute the vacuum energy with such a trivial coordinate, we use the 
\emph{interface formalism}~\cite{Graham:2001dy}, which gives the quantum 
energy per unit length in terms of the two--dimensional spectrum in the 
plane perpendicular to the string. This formulation accounts for the 
integration over the longitudinal momentum $p_z$ using sum rules for the 
scattering data~\cite{Graham:2001iv,Puff:1975zz}
to cope with the associated ultra--violet divergences.
It then remains to solve the scattering problem for the Hamiltonian
in the plane perpendicular to the string.

Although we are thus left with a seemingly well--defined two--dimensional
Dirac problem, the spectral method cannot be readily applied to compute
the vacuum energy, because the long range of the string gauge field prevents
us from setting up a well--defined scattering problem. There are  two ways to
circumvent this problem: As motivated by the study of quantum effects for 
QED flux tubes~\cite{Graham:2004jb}, a \emph{return string} was introduced in 
ref.~\cite{Weigel:2009wi} to unwind the gauge field at a large distance from 
the string core. The assumption was that the energy of the return string
is small when the unwinding is done smoothly enough and, in particular, that
the associated energy density can be well separated from the proper string
core contribution. Although these assumptions could be verified, the
necessity to repeat the (expensive) calculation of the vacuum energy with
varying return string positions to identify the core
contribution made the return string method inefficient for actual calculations.

An easier method was devised in ref.~\cite{Weigel:2010pf}. It is based on the simple
observation that the Dirac spectrum is gauge invariant, {\it i.e.} a local isospin
rotation can be used to unwind the string gauge field at spatial infinity
at the price of strong singularities in the origin (\emph{singular gauge}),
or to make the gauge field regular at the origin at the price of long--ranged fields
at spatial infinity (\emph{regular gauge}). The solution is to combine the
good features of both gauges by means of a local gauge rotation that looks
singular for large distances and regular for small distances. Thus, we make a
local gauge rotation on the Dirac Hamiltonian $H \to U^\dagger\,H\,U$ with
\begin{equation}
U=P_L {\rm exp}\,\Big(i\,\xi(\rho)\,\vek{n}\cdot\Vek{\tau}\Big)+P_R
\qquad {\rm with}\qquad
\vek{n}=\begin{pmatrix}
{\rm cos}(n\varphi) \cr -{\rm sin}(n\varphi) \cr 0
\end{pmatrix} \,.
\label{gaugetf}
\end{equation}
Here $\xi(\rho)$ is an arbitrary radial function that defines a
set of gauge transformations. Note that $\xi=0$ gives back
the original regular Hamiltonian, while $\xi=\xi_1$ together with
$f_H\equiv f_G\equiv1$ at large distances yields the return string
configuration considered in ref.~\cite{Weigel:2009wi}. Thus the interpolation
between regular and singular behavior is accomplished
by the boundary conditions  $\xi(0)=0$ and $\xi(\infty)=\xi_1$.
The transformed Dirac Hamiltonian becomes
\begin{eqnarray}
H&=&-i\begin{pmatrix}0 & \Vek{\sigma}\cdot\rhounit \cr
\Vek{\sigma}\cdot\rhounit & 0\end{pmatrix} \partial_\rho
-\frac{i}{\rho}\begin{pmatrix}0 & \Vek{\sigma}\cdot\phiunit \cr
\Vek{\sigma}\cdot\phiunit & 0\end{pmatrix} \partial_\varphi
+H_{\rm int}\,, \nonumber \\[4mm]
H_{\rm int}&=&
m f_H\left[{\rm cos}(\Delta)\begin{pmatrix} \ID & 0 \cr 0 &-\ID\end{pmatrix}
+i\,{\rm sin}(\Delta)\begin{pmatrix}0 & \ID \cr -\ID & 0\end{pmatrix}
\vek{n}\cdot\Vek{\tau}\right]
+\frac{1}{2}\frac{\partial \xi}{\partial \rho}
\begin{pmatrix}-\Vek{\sigma}\cdot\rhounit
& \Vek{\sigma}\cdot\rhounit \cr
\Vek{\sigma}\cdot\rhounit
& -\Vek{\sigma}\cdot\rhounit\end{pmatrix}\vek{n}\cdot\Vek{\tau}
\nonumber \\[4mm]
&&
+\frac{n}{2\rho}\, \begin{pmatrix}-\Vek{\sigma}\cdot\phiunit
& \Vek{\sigma}\cdot\phiunit \cr
\Vek{\sigma}\cdot\phiunit
& -\Vek{\sigma}\cdot\phiunit\end{pmatrix}
\Big[f_G\,{\rm sin}(\Delta)I_G(\Delta)+(f_G-1)\,{\rm sin}(\xi)I_G(-\xi)\Big]\,.
\label{eqDirac}
\end{eqnarray}
The new gauge function $\xi(\rho)$ is hidden in the difference
$\Delta(\rho)  \equiv \xi_1 - \xi(\rho)$ which appears both explicitly and
as the argument of the space--dependent weak isospin matrix
\begin{equation}
I_G(x)=\begin{pmatrix}-{\rm sin}(x) & -i\,{\rm cos}(x)\,{\rm e}^{in\varphi} \\[2mm]
i\,{\rm cos}(x)\,{\rm e}^{-in\varphi} & {\rm sin}(x) \end{pmatrix} \,.
\label{defIG}
\end{equation}
All explicit matrices in eq.~(\ref{eqDirac}) act in spinor space.
Together with the boundary conditions for the string profiles $f_G$ and $f_H$,
eq.~(\ref{eqDirac})  defines a well--behaved scattering problem
for which a scattering matrix and, more generally, a Jost function can be
straightforwardly computed. Moreover, the Born series to these scattering data
can be constructed simply by iterating $H_{\rm int}$.

We will renormalize the calculation by subtracting orders of
the Born series and adding these contributions back as the
corresponding Feynman diagrams. It should be mentioned that although
the Jost function is gauge invariant, neither the Born series nor the individual
Feynman diagrams associated with eq.~(\ref{eqDirac}) is gauge invariant,
and so the Born subtracted phase shifts or Jost functions will also depend
on the gauge. That is, these quantities are functionals of $\xi(\rho)$.
However, the gauge--dependent terms subtracted from the phase
shifts correspond exactly to the gauge--dependent finite parts in the
Feynman diagrams, while the counterterms, which
parameterize the ultraviolet singularities, are gauge--independent. The net
effect is that individual pieces of the spectral approach to the vacuum
energy will be gauge--dependent, but the combined expression is not.

The formulation in eq.~(\ref{eqDirac}) describes the physical string without the need
to introduce artificial return strings to unwind the topology. In particular,
the (tedious) separation of the return string contribution from the bound
state spectrum of the physical string is no longer required. Moreover, the
gauge function $\xi(\rho)$ can be taken to have support at a moderate distance,
so that there is no need for non--trivial fields at very large radii and, as a
consequence, the need for extremely large angular momenta is avoided.

To solve the scattering problem for the Hamiltonian, eq.~(\ref{eqDirac}), in 
two space dimensions, we first introduce grand--spin states to take care of
the angular dependence. For a fixed angular momentum $\ell$, there are
four grand--spin states, characterized by the quantum numbers
$\pm \sfrac{1}{2}$ for spin $S$ and isospin $I$,
\begin{equation}
\begin{array}{ll}
\langle \varphi;SI|\ell + +\rangle=
{\rm e}^{i(\ell+n)\varphi}\,
\begin{pmatrix}1 \cr 0 \end{pmatrix}_S
\otimes\begin{pmatrix}1 \cr 0 \end{pmatrix}_I & \qquad
\langle \varphi;SI|\ell + -\rangle =
-i\,{\rm e}^{i\ell\varphi}\,
\begin{pmatrix}1 \cr 0 \end{pmatrix}_S
\otimes\begin{pmatrix}0 \cr 1 \end{pmatrix}_I\cr\cr
\langle \varphi;SI|\ell - +\rangle =
i\,{\rm e}^{i(\ell+n+1)\varphi}\,
\begin{pmatrix} 0 \cr 1\end{pmatrix}_S
\otimes\begin{pmatrix}1 \cr 0 \end{pmatrix}_I & \qquad
\langle \varphi;SI|\ell - -\rangle =
{\rm e}^{i(\ell+1)\varphi}\,
\begin{pmatrix}0 \cr 1\end{pmatrix}_S
\otimes\begin{pmatrix}0 \cr 1 \end{pmatrix}_I\,.
\end{array}
\label{eq:GSstates}
\end{equation}
The angular dependence is thus separated from the radial dependence by the
{\it ansatz}
\begin{equation}
\Psi_{\ell}(\rho,\varphi) = \sum_{s,j = \pm \sfrac{1}{2}}
\Big(\langle \rho\,|\, \langle\,\varphi\,;\,S\,I\,| \Big)\,
|\epsilon \,\ell\,s\,j\,\rangle \,.
\label{GSansatz}
\end{equation}
For each value of the angular momentum $\ell$, this decomposition
turns the Dirac equation 
\begin{equation}
H\Psi=\epsilon\Psi\,,
\label{eqDirac1}
\end{equation}
with the Hamiltonian given in eq.~(\ref{eqDirac}), into a $8\times 8$
system of ordinary first order differential equation for the radial
functions in the spinor states
\begin{equation}
\begin{array}{ll}
\langle \rho |\epsilon\,\ell++\rangle =
\begin{pmatrix}f_1(\rho)|\ell + +\rangle \cr
g_1(\rho)|\ell - +\rangle \end{pmatrix} & \qquad
\langle \rho |\epsilon\,\ell+-\rangle =
\begin{pmatrix}f_2(\rho)|\ell + -\rangle \cr
g_2(\rho)|\ell - -\rangle \end{pmatrix}
\cr \cr
\langle \rho |\epsilon\,\ell-+\rangle =
\begin{pmatrix}f_3(\rho)|\ell - +\rangle \cr
g_3(\rho)|\ell + +\rangle \end{pmatrix}& \qquad
\langle \rho |\epsilon\,\ell--\rangle =
\begin{pmatrix}f_4(\rho)|\ell - -\rangle \cr
g_4(\rho)|\ell + -\rangle \end{pmatrix}\,,
\end{array}
\label{eq:GSspinors}
\end{equation}
where we have suppressed the energy label ($\epsilon$) on the radial functions.
It is convenient to combine these eight functions in a vector notation
\begin{equation}
\vec{f}=\begin{pmatrix}
f_1(\rho) \cr f_2(\rho) \cr f_3(\rho) \cr f_4(\rho)
\end{pmatrix}
\qquad {\rm and} \qquad
\vec{g}=\begin{pmatrix}
g_1(\rho) \cr g_2(\rho) \cr g_3(\rho) \cr g_4(\rho)
\end{pmatrix}\,.
\label{vecnot}
\end{equation}
In terms of these vectors, the static Dirac equation in each angular momentum
channel takes the form of two coupled real $4 \times 4$ systems,
\begin{eqnarray}
(\epsilon-m)\,\vec{f}&=&V_{uu}\,\vec{f}+\left(-CD_u+V_{ud}\right)\,\vec{g} \nonumber\\[4mm]
(\epsilon+m)\,\vec{g}&=&\left(CD_d+V_{du}\right)\,\vec{f}+V_{dd}\,\vec{g}
\label{DiracMatrix}\,.
\end{eqnarray}
The $4\times4$ matrix $C={\rm diag}(-1,-1,+1,+1)$ is constant while
$D_u$ and $D_d$ contain the radial derivative operator as well as the
angular barrier terms. The coupling to the background profiles of the
boson fields emerges via the matrices $V_{ij}$. Detailed expressions
for $D_u$, $D_d$ and $V_{ij}$ are listed in appendix~\ref{appB}.
The ODE system eq.~(\ref{DiracMatrix}) is the basis of the spectral approach
to the string problem.

For the gauge profile $\xi(\rho)$, any smooth function
with $\xi(0) = 0$ and $\xi(\infty) = \xi_1$ will do. For simplicity
and to avoid possible singularities at $\rho\to 0$, we choose
again a Gau{\ss}ian profile
\begin{equation}
\xi(\rho) = \xi_1\,\Big[1 - \exp\left(-\rho^2 / w_\xi^2\right)\Big]
\label{xi}
\end{equation}
with a new width parameter $w_\xi$. As explained earlier, the
scattering matrix without Born subtractions and the complete quantum energy
should be independent of the choice of gauge and thus
independent of the width parameter $w_\xi$. This has been verified
numerically to a fairly high precision \cite{Weigel:2010pf}.

\section{Spectral method}
\label{sec:4}
\noindent
In this section, we present the details of our approach
to compute the fermion contribution to the vacuum energy of the string.
To make the exposition clearer, we have moved overly complicated expressions
and all technical derivations to the appendices. However, the complete
method is still quite involved due to the many contributions that enter. 
We will continue the discussion of the variational approach for charged strings
in section \ref{sec:5} and present numerical results in section \ref{sec:6}.

The calculation of the fermion quantum energy is based on the Dirac 
equation~(\ref{eqDirac1}). From the solutions to this equation we infer
a number of distinct contributions to the energy of the string,
\begin{equation}
E_f = E_{\delta} + E_{\rm FD} + E_{\rm b}
\label{energy:quantum}
\end{equation}
In physical terms, these three contributions are
\begin{enumerate}
\item[$E_\delta$:] the \emph{non--perturbative} vacuum polarization due
to the string background, with the divergent low--order Feynman
diagrams taken out by subtracting leading terms in the Born
expansion. This piece also includes the bound state contribution  to the 
fermion determinant; 
\item[$E_{\rm FD}$:] the \emph{perturbative} contribution of the
low--order Feynman diagrams to the vacuum polarization energy,
combined with the counterterms for proper renormalization. This compensates for 
the part that has been taken out of
$E_\delta$ by means of the corresponding Born expansion;
\item[$E_{\rm b}$:] the binding energy due to the single particle
bound states that are \emph{explicitly} occupied
to give the string a fermion charge $Q$. More precisely,
$E_{\rm b}=\Big[\sum\limits_{\rm occ~bs}\epsilon_i\Big]-Qm$ measures the energy
of the populated levels relative to the same number of free fermions.
We will describe this contribution in the next section.
\end{enumerate}
Each of these pieces is separately finite; the first two terms are
\emph{not} gauge invariant, but their sum is, and so is $E_{\rm b}$.

In this section we focus on the renormalized vacuum polarization energy 
\begin{equation}
E_q=E_{\delta} + E_{\rm FD}\,.
\label{defEq}
\end{equation}
Potential ambiguities in $E_q$ that could originate from the ultra--violet divergences are
fully removed  by the identification of terms in the Born series with Feynman diagrams.
The most important feature of $E_q$ is the possibility to impose renormalization conditions
from the  perturbative sector ($\overline{\rm MS}$ or \emph{on--shell}), although the
calculation is completely non--perturbative, including all orders in $H_{\rm int}$. We 
then combine $E_q$ with the classical energy $E_{\rm cl}$ required to form the bosonic
string background. Quite generally, we expect $E_{\rm cl}+E_q>0$ once quantum 
fluctuations are included, since  otherwise we would have an instability of the true 
vacuum to cosmic string condensation, which should obviously not happen.

In the following subsections, we will give brief accounts for each contribution
to eq.~(\ref{defEq}). 
More details can be found in the appendices.

\subsection{Jost function and Born subtractions}

The Born--subtracted vacuum polarization energy $E_q$ has contributions from 
bound and scattering states.  These two contributions are combined in the 
Jost function for imaginary momenta~\cite{Bordag:1994jz,Graham:2009zz}. 
To compute $E_q$ it is therefore sufficient to solve the scattering  problem as 
in ref.~\cite{Weigel:2009wi}: For every energy $|\epsilon|>m$, the fermion 
system eq.~(\ref{DiracMatrix}) has eight real linear independent solutions 
$(\vec{f},\vec{g})$. In the case without a string background, these solutions 
are Bessel functions of integer order with the argument $z=k\rho$, where 
$k=\sqrt{\epsilon^2-m^2}\ge0$. Instead of taking the (real) regular and singular
Bessel functions $J_\nu(z)$ and $Y_\nu(z)$, respectively, we can formally let
$(\vec{f},\vec{g})$ have complex coefficients and take Hankel function
solutions instead. In this case, both the real and imaginary
parts of $(\vec{f},\vec{g})$ are (linearly independent) solutions, or
equivalently $(\vec{f},\vec{g})$ and their complex
conjugates are independent solutions.

To describe the coupled channel scattering problem, it is convenient to put the
four free linearly independent complex solutions for $\vec{f}$
and $\vec{g}$ onto the diagonal of two $4\times4$ matrices
\begin{eqnarray}
\mathcal{H}_u&=&\mbox{diag}\left(
H^{(1)}_{\ell+n}(k\rho),H^{(1)}_{\ell}(k\rho),
H^{(1)}_{\ell+n+1}(k\rho),H^{(1)}_{\ell+1}(k\rho)\right) \\[2mm]
\mathcal{H}_d&=&\mbox{diag}\left(
H^{(1)}_{\ell+n+1}(k\rho),H^{(1)}_{\ell+1}(k\rho),
H^{(1)}_{\ell+n}(k\rho),H^{(1)}_{\ell}(k\rho)\right)\,,
\label{Smat2}
\end{eqnarray}
which describe out--going asymptotic fields since
\begin{equation}
H_\nu^{(1)}(z)=J_\nu(z)+iY_\nu(z)\longrightarrow
\sqrt\frac{2}{\pi z} \,
{\rm e}^{i\left(z-\frac{\nu}{2}\pi-\frac{1}{2}\pi\right)}\,,
\label{outgoing}
\end{equation}
as $z\to\infty$.
With this notation, the $j^{\rm th}$ linear independent solution is
\[
(\vec{f})_j = \left[\mathcal{H}_u\right]_j\,,\qquad\quad
(\vec{g})_j = \kappa\cdot\left[\mathcal{H}_d\right]_j
\]
where $[\mathcal{H}]_j$ denotes the $j^{\rm th}$ row  of the matrix $\mathcal{H}$.
For convenience we omit the orbital angular momentum index $\ell$.
By construction, the complex conjugate matrices, $\mathcal{H}_{u,d}^\ast$
describe incoming spherical waves. Furthermore, we have defined the relative 
weight of upper and lower Dirac components as
\begin{equation}
\kappa \equiv \frac{k}{\epsilon+m} = \frac{\epsilon-m}{k}\,.
\end{equation}
For later analytic continuation we must ensure that the phase of the
Jost function is odd for real momenta under $k\to-k$, which requires the branch
cut structure of the square root be defined using either of the two expressions
listed above.

To describe the coupling of the four channels in the actual scattering
problem, it is convenient to put the linearly independent solutions 
for $\vec{f}$ and $\vec{g}$ again in the rows of a $4\times 4$ matrix, 
and factor out the free part to get simple Jost boundary conditions,
\begin{equation}
\begin{array}{r@{\,\,\,\,\longrightarrow\,\,\,\,}ll@{\qquad\mbox{and}\qquad}ll}
(\vec{f})_j & & \displaystyle\left[\mathcal{F} \cdot \mathcal{H}_u\right]_j &
& \displaystyle\left[\mathcal{F}^* \cdot \mathcal{H}^\ast_u\right]_j
\\[3mm]
(\vec{g})_j &  \kappa&\displaystyle\left[\mathcal{G} \cdot \mathcal{H}_d\right]_j &
\kappa& \displaystyle\left[\mathcal{G}^* \cdot \mathcal{H}^\ast_d\right]_j \,.
\end{array}
\label{Smat1}
\end{equation}
We substitute these {\it ans\"atze} into the Dirac equation,
eq.~(\ref{DiracMatrix}),
and find first order differential equations for the matrices $\mathcal{F}$
and $\mathcal{G}$. This is explicitly carried out in appendix~\ref{appB1}.
The solutions to eq.~(\ref{deqforGF}) with the Jost boundary conditions
\begin{equation}
\lim_{\rho\to\infty} \mathcal{F}(\rho)
= \lim_{\rho\to\infty}\mathcal{G}(\rho) = \ID
\end{equation}
define Jost solutions to the initial Dirac problem via the
representation in eq.~(\ref{Smat1}). The physical scattering solution is the
linear combination which at large distances is the superposition of
incoming and outgoing free spherical waves and obeys the regularity condition 
at the origin.  The relative weight of the incoming and outgoing waves defines
the scattering matrix $\mathcal{S}$. Hence the physical scattering solution for 
the $\mathcal{F}$--type (upper) components reads
\begin{equation}
\Psi = \mathcal{F}^\ast\cdot \mathcal{H}_u^\ast +
(\mathcal{F}\cdot \mathcal{H}_u) \cdot \mathcal{S}.
\label{scatsol}
\end{equation}
The corresponding $\mathcal{G}$--type (lower) components 
are obtained by replacing
$\mathcal{F} \to \mathcal{G}$ and $\mathcal{H}_u \to \kappa
\mathcal{H}_d$.
The physical scattering solution must be regular at the origin
$\rho=0$. From this condition we extract the scattering matrix in one
of two equivalent ways
\begin{eqnarray}
\mathcal{S}&=&-\lim_{\rho\to0}\,
\mathcal{H}_u^{-1}\cdot\mathcal{F}^{-1}\cdot
\mathcal{F}^*\cdot\mathcal{H}_u^\ast \\[2mm]
\mathcal{S}&=&-\lim_{\rho\to0}\,
\mathcal{H}_d^{-1}\cdot\mathcal{G}^{-1}\cdot
\mathcal{G}^*\cdot\mathcal{H}_d^\ast \,.
\label{Smat3}
\end{eqnarray}
The phase convention in
eq.~(\ref{scatsol}) is chosen to reproduce $\mathcal{S} = \ID$ for the
non--interacting case which has $\mathcal{F}^{(0)}=\mathcal{G}^{(0)}=\ID$.
The equality of the two representations from the system of coupled differential equations
is a good check on our numerics, as is the requirement that $\mathcal{S}$ be unitary.

It should finally be noted that all the interaction matrices eq.~(\ref{intmatrix})
are \emph{linear} in the background profiles eq.~(\ref{alphas}), so that the
ODE system for the Born approximation can simply be obtained by iteration with
the {\it ansatz}
\begin{equation}
\mathcal{F}_{\rm Born}(\rho)=\sum_{i=0}^\infty \mathcal{F}^{(i)}(\rho)
\qquad {\rm and} \qquad
\mathcal{G}_{\rm Born}(\rho)=\sum_{i=0}^\infty
\mathcal{G}^{(i)}(\rho)\,,
\label{defBorn}
\end{equation}
where the superscript denotes the order of the interaction Hamiltonian $H_{\rm int}$ in
eq.~(\ref{eqDirac}). The zeroth order solutions
are $\mathcal{F}^{(0)}(\rho)=\mathcal{G}^{(0)}(\rho)=\ID$ and
all subsequent contributions are subject to the boundary conditions
\[
\lim_{\rho\to\infty} \mathcal{F}^{(i)}(\rho)
=\lim_{\rho\to\infty} \mathcal{G}^{(i)}(\rho) = 0\,,
 \qquad\quad i = 1,2,3\ldots
\]
The explicit form of the iterated system of differential equations
for the Born approximations of order $i=1$ and $i=2$ can be found in
appendix~\ref{appB}. Though the $i=3$ and $i=4$ orders also yield
divergences, we do not discuss them explicitly because we employ a
numerically less costly method to  handle these logarithmic
divergences, as described below.

\subsection{Interface formalism}
\noindent
The $4\times4$ scattering matrix $\mathcal{S}$ derived in the last
subsection yields the four eigenphase shifts and thus the shift in the
two--dimensional density of states~\cite{Graham:2002xq}
\begin{equation}
\rho_\ell(k) - \rho_\ell^{(0)}(k)
= \frac{1}{\pi}\,\sum_{c=1}^4\,\frac{d\delta_{\ell,c}}{d k}
= \frac{i}{2\pi}\,\frac{d}{dk}\,\ln\mathrm{det}\,\mathcal{S}_\ell(k) \,,
\label{changedensity}
\end{equation}
where the sum runs over the four scattering channels for a given grand
spin channel, which we label by the associated orbital angular momentum
$\ell$. To turn this two--dimensional density into a three--dimensional energy
(or energy per unit length of the string), we have to deal with the trivial
dynamics along the string symmetry axis of the string. This is a
typical application of the \emph{interface formalism} developed in
ref.~\cite{Graham:2001dy}. The modifications of the usual spectral method
are simple:
\begin{enumerate}
\item The integration over the momentum conjugate to the coordinate of 
translational invariance remains finite due to sum rules for scattering
data~\cite{Graham:2001iv,Puff:1975zz} that are generalizations of Levinson's 
theorem.
\item  
When integrating over momentum $k$, the density, eq.~(\ref{changedensity}),
must be multiplied by a kinematic factor that differs
from the usual one--particle energy $\epsilon= \sqrt{m^2 + k^2}$.
\item More Born subtractions are required to make the momentum integral
and angular momentum sum convergent. This corresponds to the larger
number of divergent Feynman diagrams in three dimensions.
\end{enumerate}
Then the interface formula for the vacuum polarization energy per unit 
length of the string is
\begin{eqnarray}
E_\delta^{(N)}&=&\frac{1}{4\pi}\sum_{\ell=-n}^\infty \Bigg\{D_\ell
\int_0^\infty \frac{dk}{\pi}
\left[(k^2+m^2) {\rm ln}\left(\frac{k^2+m^2}{\mu^2}\right)-k^2\right]\,
\frac{d}{dk}\left[\delta_\ell(k)\right]_N \cr
&& \hspace{2cm}
+\sum_j\left[\left(\epsilon_{j,\ell}\right)^2
{\rm ln}\frac{\left(\epsilon_{j,\ell}\right)^2}{\mu^2}
-\left(\epsilon_{j,\ell}\right)^2+m^2\right]\Bigg\}\,,
\label{interface:real}
\end{eqnarray}
where the notation $[\cdots]_N$ refers to the quantity in
the brackets with its first $N$ terms of the Born series
subtracted. For the string problem in three space dimensions, we need
$N \ge 4$ to ensure convergence of the momentum
integral although, as described below, we will use a different
subtraction in place of the $N=3$ and $N=4$ cases.
Here, $\epsilon_{j,\ell}$ gives the energy of the $j^{\rm th}$ bound state
in the angular momentum channel $\ell$, and $D_\ell$ is the degeneracy in
that channel. For the string background, we have
\begin{equation}
D_\ell=\begin{cases}
1\,, \quad & \ell=-n \cr
2\,, \quad & \ell > -n\,,
\end{cases}
\label{degeneracy}
\end{equation}
where $n=1$ is the Higgs winding number introduced in the string configuration
of eqs.~(\ref{string_gauge}) and~(\ref{string_higgs}).
The renormalization scale $\mu$ emerged from the integration 
under item 1. It cancels due to the same sum rules. For convenience 
we usually set $\mu=m$.
The phase shifts can be extracted from the scattering matrix or,
equivalently, from the Jost--like matrices $\mathcal{F}$ and $\mathcal{G}$
introduced in the last subsection,
\begin{equation}
\delta_\ell(k)=\frac{1}{i}\,{\rm ln} \,{\rm det}\,
\lim_{\rho\to0}\mathcal{F}_\ell(\rho,k)^{-1}\mathcal{F}^\ast_\ell(\rho,k)
=\frac{1}{i}\,{\rm ln} \,{\rm det}\,
\lim_{\rho\to0}\mathcal{G}_\ell(\rho,k)^{-1}
\mathcal{G}^\ast_\ell(\rho,k)\,,
\label{delta}
\end{equation}
where we have restored all the arguments. In deriving eq.~(\ref{delta}) from
eq.~(\ref{Smat3}) we have used the cyclic property of the trace and the
fact that as $\rho\to0$ the Hankel functions are dominated by their
imaginary parts.

As indicated in the previous subsection, it is convenient to evaluate 
the expression (\ref{interface:real}) in the complex $k$--plane because 
after rotating to the imaginary axis the explicit bound state contribution
is automatically canceled by the pole contribution from Cauchy's theorem,
leaving only a single integral along the cut on the positive imaginary 
axis~\cite{Bordag:1994jz,Graham:2009zz}.
There is another important technical reason to rotate to imaginary
momentum. We need to sum over angular momentum $\ell$ and integrate
over radial momentum $k$ after subtracting sufficiently many terms
of the Born series. This procedure is numerically cumbersome
because these functions oscillate in $k$, which can make it impossible to
exchange the sum and integral~\cite{Schroder:2007xk} because they are not
absolutely convergent. This obstacle is also avoided
by analytically continuing to imaginary momenta $t=ik$ and performing
the integrals in the complex plane along the branch cut
$t>m$~\cite{Schroder:2007xk}.

The analytic continuation for the Dirac equation is conceptually 
different from the well--studied Schr\"odinger case because 
$\epsilon=\pm\sqrt{k^2+m^2}$ causes the complex
momentum plane to have two sheets. So on the real axis we have to pick
one sign, continue to complex momenta and compute the Jost function on
the imaginary axis. This procedure must then be repeated for the other
sign and then all discontinuities must be collected at the end. In the
present problem we are fortunate because the solutions to the Dirac
equation exhibit charge conjugation symmetry along the real axis.
Therefore ${\rm det}(\mathcal{S})$ does not change under 
$\epsilon\to-\epsilon$ and there is no additional discontinuity in 
the Jost function choosing either sign. Moreover,
the Jost function is real on the imaginary axis, as in the Schr\"odinger
problem. However, the way this comes about in the string problem
requires us to be careful when constructing the Jost function for
complex momenta. This procedure is described
in appendix~\ref{appB} and results in the replacement of the phase shift
$\delta(k)$ (and its Born expansion) by $\nu(t)$, the (modified) logarithmic
Jost function for imaginary momentum. For this to work it is
essential to have $\kappa$ odd under sign reflection of real $k$.
The resulting Jost function itself is a continuous function in the upper
complex momentum plane and the branch cuts in the Dirac equation do not 
carry over to $\nu(t)$. The only discontinuity arises from the logarithm under
the integral in eq.~(\ref{interface:real}), which is $2\pi$. Finally
an integration by parts yields a simple expression for
the Born subtracted vacuum polarization energy,
\begin{equation}
E_\delta^{(N)} = - \frac{1}{2 \pi} \int\limits_{m}^\infty
dt\,t \sum_{\ell=-n}^\infty D_\ell \left[\nu_\ell(t)\right]_N\,.
\label{interface::imag0}
\end{equation}
Here we have interchanged the integral with the angular momentum sum, which is
possible on the imaginary axis~\cite{Schroder:2007xk}. After a 
final change of variables $t\to\tau=\sqrt{t^2-m^2}$, we obtain eventually
\begin{equation}
E_\delta^{(N)}=-\frac{1}{2\pi} \int_0^\infty d\tau\, \tau
\sum_{\ell=-n}^\infty D_\ell \left[\nu_\ell(\sqrt{\tau^2+m^2})\right]_N \,.
\label{interface:imag}
\end{equation}
Eq.~(\ref{interface:imag}) is our master formula for the phase shift
contribution to the vacuum polarization energy per unit length of the
string.

\subsection{Feynman diagrams}
\noindent
The Born subtractions in the integrand of eq.~(\ref{interface:imag}) must be added
back in as Feynman diagrams. The latter are most easily derived by expanding the
fermion determinant representation of the (unrenormalized) vacuum polarization energy,
\begin{equation}
\mathcal{A} \equiv - TL\,E_{\rm q} = -i \,\ln\,\det\Big[i \dslash-m + H_I \Big]\,.
\end{equation}
Both the time interval $T$ and the length $L$ of the string
factorize as $T\to\infty$ and $L\to\infty$ because the string
background is static and translationally invariant. The interaction
part of the Dirac operator can be separated in various spin structures,
\begin{equation}
H_I=L_\mu\gamma^\mu P_L +h +ip\gamma_5
\label{hint1}
\end{equation}
where the fields $L_\mu$, $h$ and $p$ are isospin $2\times 2$ matrices,
\begin{equation}
\begin{array}{ll}
L_0=0\,,\quad
& \vek{L}(\rho)=2\,\ar(\rho)\, I_P(\varphi)\,\rhounit
+2\Big[\ag(\rho)\, I_G(\xi_1 - \xi(\rho))+\ax(\rho)\, I_G(-\xi(\rho))\Big]
\,\phiunit\,,
\\[2mm]
h(\rho)=-\alpha_H(\rho)\,\ID\,,\quad
& p(\rho) = -\alpha_P(\rho)\, I_P(\varphi) \,.
\end{array}
\label{hint2}
\end{equation}
The isospin matrices $I_G$, defined in eq.~(\ref{defIG}), and
\begin{equation}
I_P(\varphi) = \vek{n}\cdot \Vek{\tau} =
\begin{pmatrix} 0 & {\rm e}^{in\varphi} \cr
{\rm e}^{-in\varphi} & 0 \end{pmatrix}
\label{Ip}
\end{equation}
contain the entire dependence on the azimuthal angle $\varphi$.
The profile functions $f_H(\rho)$, $f_G(\rho)$ and $\xi(\rho)$, {\it cf.}
eqs.~(\ref{eqn:profile}) and~(\ref{xi}), determine the radial behavior of the
coefficient functions $\ar(\rho)$, $\alpha_P(\rho)$ and $\ax(\rho)$. They
are explicitly listed in eq.~(\ref{alphas}) of the appendix.
The Feynman series for the effective fermion action (determinant) is now
\begin{equation}
\mathcal{A} \equiv
- T L \,E_{\rm q} = -i\,\ln \,\det \left(i \dslash -m \right) + \sum_{N=1}^\infty
\frac{(-1)^N}{N}\,\mathrm{Tr}\,\Big[\left(i \dslash - m\right)^{-1}\,H_I\Big]^N
\label{Feynmanseries}
\end{equation}
where the first term corresponds to the free vacuum energy without a
string background. It is automatically removed in the spectral method
by the difference in eq.~(\ref{changedensity}).
For fermions in three dimension, all diagrams up through $N=4$ are
divergent and thus subject to renormalization. While the calculation
of the corresponding Born subtractions up to fourth order is not
particularly hard, the evaluation of the higher order Feynman diagrams
with up to four nested Feynman parameter integrals and an equal number
of Fourier transformations of the string background is very cumbersome.
A better approach is the so--called \emph{fake boson method} introduced
in ref.~\cite{Farhi:2001kh}, which we will describe next.

\subsection{Fake boson approach and renormalization}
\label{boson:fake}

The first and second order fermion Feynman diagrams contain both
quadratic and subleading linear and logarithmic ultra--violet divergences,
so that a precise identification of the terms in the Born
expansion with the Feynman diagrams must be made separately for each
term in the angular momentum sum. On the other hand, the third and fourth
order fermion Feynman diagrams only cause logarithmic divergences, which are
much easier to cope with, because the sum
\begin{equation}
s(t) \equiv \sum_\ell D_\ell\left[\nu_\ell(t)\right]_2
\label{integrand2}
\end{equation}
is finite. However, after multiplication by $t$, the integral in
eq.~(\ref{interface::imag0}) is logarithmicly divergent. So instead of subtracting
the complete third and fourth order terms in $H_I$ from the sum in eq.~(\ref{integrand2}),
it is sufficient to  just subtract \emph{any} function $\Delta s(t)$ of momentum
with the same ultra-violet behavior, provided that the following conditions
are met:
\begin{enumerate}
\item the subtraction $\Delta s(t)$ should have the same analytic properties 
with respect to complex momentum arguments;
\item  formally its contribution to the vacuum polarization should be
identifiable as a Feynman diagram that can be combined with the 
available counterterms
\end{enumerate}
\begin{equation}
\mathcal{L}_{\rm ct} = c_1\,\mathrm{tr}\big(G^{\mu\nu}\,G_{\mu\nu}\big) +
c_2 \,\mathrm{tr}\,\Big[\,\big(D^\mu \Phi\big)^\dagger\,D_\mu \Phi\Big] +
c_3 \Big[ \mathrm{tr}\big(\Phi^\dagger\Phi\big) - 2 v^2 \Big] +
c_4 \,\Big[\mathrm{tr}\big(\Phi^\dagger\Phi\big)-2 v^2 \Big]^2\quad{}
\label{Lcounter}
\end{equation}
to cancel all ultra--violet divergences. The perfect candidate is the
second order contribution from a boson scattering off a radially symmetric
potential $V(\rho)$. From the properties of the (bosonic) scattering problem
\cite{Graham:2002xq}, we know that its Jost function has the required
analytical properties and its contribution to the  vacuum polarization energy
can be expressed as a (very simple) Feynman diagram.  It only remains to adjust
its strength to accomplish the required
subtraction.  This \emph{fake boson} scattering problem also has a partial
wave decomposition and we subtract the sum of the logarithm of the second 
order fake boson Jost function from the sum in eq.~(\ref{integrand2}).
Since the subtraction is not carried out channel by channel,
the exchange of $\ell$--sum and $t$--integral is crucial for this
approach to work.

To describe the method in detail we define $\mathcal{A}_n$ to be the
contribution of order $(H_I)^n$ in the sum in
eq.~(\ref{Feynmanseries}).
\begin{enumerate}
\item
The first order diagram $N=1$ is linear in the interaction $H_I$ and
local, including all finite parts. Thus the entire diagram is proportional to
the spacetime integral of the $c_3$--counterterm in eq.~(\ref{Lcounter}).  
We fix the corresponding counterterm by the \emph{no--tadpole} condition 
$\mathcal{A}_1\stackrel{!}{=}0$, which ensures that  the \emph{vev} of the 
Higgs field is kept at its classical value $v$.
This condition completely fixes both the divergence and the finite part in
the $c_3$--counterterm, {\it cf.} eq.~(\ref{finitecounter}).
\item The second order diagrams $N=2$ give contributions to the
various propagators whose contributions to the vacuum polarization
energy are quadratically  divergent at large momenta, for which a careful
regularization is required. Due to gauge invariance, the coefficients
$c_1$, $c_2$ and $c_4$ in the counterterms in
eq.~(\ref{Lcounter}) can unambiguously be determined by the two--point
functions that emerge at order $N=2$. Hence we do  not need to compute
the full Feynman diagrams at orders $N=3$ and $N=4$.
\item Although we do not need the full diagrams,
we do need to precisely subtract the divergences
from $\mathcal{A}_3$ and $\mathcal{A}_4$. In dimensional
regularization ($D\to4$), these logarithmicly divergent pieces read
\begin{equation}
\mathcal{A}^{\infty}_{3,4}= \pi  c_F\, TL\,
\left[i\left(\frac{\mu}{m}\right)^{4-D}
\int\frac{d^D k}{(2\pi)^D}\left(k^2-1+i\epsilon\right)^{-2}\right]\,,
\label{logdiv1}
\end{equation}
where $T$ and $L$ are the (infinite) lengths of the time and $z$--axis
intervals, respectively, and $c_F$ is a complicated integral over the
radial profile  functions, {\it cf.} eq.~(\ref{lfctcoef}). The key
observation for the implementation of the fake boson approach is that the
divergence in eq.~(\ref{logdiv1}) is also contained in the
two--point function of a simple scalar field that fluctuates in a
(fictitious) background potential $V(\rho)$. In fact, the divergence
in the second order boson diagram has the form of
eq.~(\ref{logdiv1}) with $c_F$ replaced by
\begin{equation}
c_B = \frac{1}{4} \int_0^\infty d\rho\,\rho \,V(\rho)^2\,.
\end{equation}
By properly scaling $V(\rho)$ with $\sqrt{c_F/c_B}$,
we can match the divergences from eq.~(\ref{logdiv1}). The equivalence of the
Feynman and Born expansions implies that the combination of $s(t)$ with
\begin{equation}
\Delta s(t) = \frac{c_F}{c_B} \sum_\ell \overline{D}_\ell
\overline{\nu}_\ell^{(2)}(t)
\label{fakesub}
\end{equation}
is finite when integrated according to
equation~(\ref{interface::imag0}). Here $\overline{\nu}_\ell^{(2)}(t)$
is the second order Born approximation for logarithm of the Jost
function on the imaginary axis in the fake boson
problem and the associated degeneracy factor in the
partial wave decomposition  is $\overline{D}_\ell=2-\delta_{0,\ell}$.
\item
The subtraction in eq.~(\ref{fakesub}) must be compensated by
adding the corresponding second order fake boson diagram. Since the
divergences of the fake boson and the fermion problem have been carefully
matched, the fermion counterterms from eq.~(\ref{Lcounter}) are sufficient to
render the relevant fake boson diagram finite. As a consequence, only the
renormalized fake boson diagram must be added back in, {\it cf.} appendix
\ref{app_fake}. We are then fully prepared to compute the vacuum
polarization energy in any renormalization scheme. We first 
consider the $\overline{\rm MS}$ scheme, which is defined by 
setting $\overline{c}_s=0$ in the counterterm coefficients
\begin{equation}
c_s=-i\left(\frac{\mu}{m}\right)^{4-D}
\int \frac{d^D k}{(2\pi)^D}\left(k^2-1+i\epsilon\right)^{-2}
+\overline{c}_s\,,\qquad\quad s = 1,2,4 \,.
\label{finitecoef1}
\end{equation}
In this scheme the dependence on the model parameters is simple. The 
computational advantages of first considering the $\overline{\rm MS}$ scheme will
be discussed thoroughly in section \ref{sec:6}.
\end{enumerate}
Let us summarize the result in the $\overline{\rm MS}$ scheme and
carefully describe the angular momentum sums. First we
construct the subtracted logarithmic Jost function for imaginary momenta
\begin{equation}
\nu(t)=\lim_{\ell_{\rm max}\to\infty}\sum_{\ell=-n}^{\ell_{\rm max}}
D_\ell\left[\nu_\ell(t)\right]_2
+\frac{c_F}{c_B}\,
\lim_{\overline{\ell}_{\rm max}\to\infty}
\sum_{\ell=0}^{\overline{\ell}_{\rm max}}
\overline{D}_{\ell}\,\overline{\nu}_{\ell}^{(2)}(t)\,.
\label{finalJost}
\end{equation}
From eq.~(\ref{finalJost}), we can compute the phase shift contribution to
the vacuum polarization energy,
\begin{equation}
E_\delta=-\frac{1}{2\pi} \int_0^\infty d\tau\, \tau\, \nu(\sqrt{\tau^2+m^2})\,.
\label{MASTER2}
\end{equation}
The complete vacuum polarization energy in the $\overline{\rm MS}$ scheme is
then the sum
\begin{equation}
E_{\overline{\rm MS}} = E_\delta+ \Delta E_{\rm FD} \,,
\label{VERYMASTER}
\end{equation}
where $\Delta E_{\rm FD}=\Delta E^{(2)}_{\rm FD}+\Delta E_{\rm B}$
is the sum of the renormalized values
(finite parts in $\overline{\rm MS}$) of the second order fermion and fake
boson diagram. Explicit expressions for these contributions can
be found in eqs.~(\ref{fdint0}) and~(\ref{fdbos}). As a further test
of the approach we verify numerically that $E_{\overline{\rm MS}}$
remains unchanged when the boson potential $V(\rho)$ is modified.

To make contact with the electroweak theory, it is convenient to re--adjust the
finite pieces in the counterterms such that they match the so--called
\emph{on-shell scheme}. In addition to the already implemented no--tadpole condition
that fixes $c_3$, we thus require
\begin{itemize}
\item The pole of the Higgs propagator remains at the tree level mass,
$m_h=m_h^{(0)}$, with unit residue. This fixes the coefficients $c_2$ and $c_4$
and ensures the usual one--particle interpretation of the states created by the
asymptotic Higgs field.
\item The residue of the gauge field propagator (in unitary gauge) is unity, so that
asymptotic $W$--fields create one--particle $W$--boson states.
This condition determines~$c_1$.
\end{itemize}
The position of the pole in the gauge boson propagator is then a prediction,
{\it i.e.}~the physical $W$-boson mass receives radiative corrections. These
corrections are determined by the implicit solution to eq.~(\ref{implicitmass})
presented in appendix \ref{app_onshell}.

The on--shell and $\overline{\rm MS}$ schemes are related by finite changes
of the counterterm coefficients, so that $\overline{c}_i\neq 0$ in the
on--shell scheme; explicit expressions can again be found in appendix
\ref{app_onshell}. The modification to the vacuum polarization energy
eq.~(\ref{VERYMASTER}) due to the change in renormalization scheme is then simply
the energy from the counterterm Lagrangian eq.~(\ref{Lcounter}), with the coefficients
$c_i$ replaced by their finite pieces $\overline{c}_i$ listed in eq.~(\ref{finitecounter}).
Since the counterterms are local, this modification amounts to a 
radial integral similar to the classical energy in eq.~(\ref{eq:ecl}), 
which is numerically inexpensive. Hence the vacuum polarization energy in
the on--shell scheme is
\begin{equation}
E_{\rm q}=E_{\overline{\rm MS}}+E_{\rm CT}=
E_{\delta}+\Delta E^{(2)}_{\rm FD}+\Delta E_{\rm B}+E_{\rm CT}\,.
\label{eq:evac}
\end{equation}
The explicit expression for the counterterm contribution reads
\begin{eqnarray}
E_{\rm CT}&=&2\pi\int_0^\infty \rho d\rho\,\Bigg\{
{\rm sin}^2\xi_1\, \frac{n^2}{\rho^2}
\left[\overline{c}_2\,v^2 f_H^2\left(1-f_G\right)^2
-\frac{4\overline{c}_1}{g^2}f_G^{\prime2}\right]\cr
&&\hspace{3cm}
+\overline{c}_2\, v^2 f_H^{\prime2}
-2\overline{c}_4\, v^4 \left(1-f_H^2\right)^2\Bigg\}
\label{eq:ect}
\end{eqnarray}
The counterterm coefficient $\overline{c}_3$ does not appear explicitly 
because this counterterm receives no correction in passing between
the $\overline{\rm MS}$ and on--shell schemes.

\section{Charged String}
\label{sec:5}
As already discussed in refs.~\cite{Weigel:2009wi,Weigel:2010pf}, the
fermion vacuum energy is negative for narrow strings and thus provides
some binding. However, for physically relevant model
parameters, eq.~(\ref{eq:parameters}), it is insufficient to
overcome  the large classical energy. The central mechanism for
overcoming the classical energy cost is to populate the numerous
fermion bound states that emerge in the background of the string,
which as a result assigns charge $Q$ to the string. If the energy of
equally many free fermions $Qm$ is larger than the total energy of the
string (the classical, vacuum polarization and contribution
from populated levels combined), we have succeeded in constructing a
stable charged string. Quantitatively, this requirement corresponds to
$E_{\rm cl}+\mbox{$E_f$}\le0$,  {\it cf.} eqs.~(\ref{eq:ecl})
and~(\ref{energy:quantum}). It prevents the direct decay into
fermions and only leaves charge non--conserving decay channels,
where the decay rate is heavily suppressed due to the sphaleron
barrier. The direct decay into lighter fermion doublets is also
suppressed, since we do not have flavor mixing in our model.
To carry out this procedure, we first need
to find the bound state energies in the string background.

\subsection{Bound states and box diagonalization}
\noindent
Since our system is translation invariant in the $z$ direction, we begin
by finding Dirac bound states, $\epsilon_i$, of the two--dimensional
problem, eq.~(\ref{eqDirac1}).  Each bound state we find will then
correspond to a family of bound states in the three--dimensional problem,
indexed by the transverse momentum~$p_z$. 

We carry out the two--dimensional bound state calculation by putting 
the string in a large
cylindrical box of radius $R \gg m^{-1}$ and imposing the boundary condition
that no net flux runs through the surface of the cylinder. This
boundary condition discretizes the possible radial momenta in each
angular momentum channel through the roots of certain Bessel
functions, {\it cf.} eq.~(\ref{discret1}) for the case of unit winding, $n=1$.
We can thus take a countable set of grand spin solutions,
eq.~(\ref{eq:GSstates}), to the free Dirac equation and express the
fully interacting string Hamiltonian as an infinite matrix in this
basis. The relevant matrix elements are again presented in appendix
\ref{appA}. Upon truncating the set of free solutions by including an
effective UV cutoff $\Lambda$ on the discrete momenta, we are thus
left with a large matrix diagonalization in each grand spin channel to
determine the fermion eigenstates in the string background. Typical matrix 
sizes are $(1600 \times 1600)$ including Dirac indices We then
find the energy eigenvalues numerically by diagonalization.
In the finite box, of course, \emph{all} energy levels are discrete
and there are no continuum states. In the limit $R \to \infty$ and
$\Lambda \to \infty$, the highest energy levels in the quasi--continuum
will still fluctuate considerably, but the low--lying bound state
spectrum of states with energy smaller than $m$, which become bound
states in the $R\to \infty$ limit, should remain stable. This was indeed
observed for moderate values, $\Lambda \approx 8m$ and $R \approx 75/m$.

It should be noted that bound states occur predominantly in the lower angular
momentum channels, as we would expect since the higher channels
contain an increasingly large centrifugal barrier.
Depending on the width of the background profile, we see bound states
in as many as $10$  channels, or only in the single channel
$\ell=-n=-1$, which is the channel that contains an exact zero mode
for $\xi_1 = \sfrac{\pi}{2}$.

A good numerical test on our diagonalization procedure is the gauge invariance
of the Dirac Hamiltonian and thus of the bound state spectrum. In our specific
case, this means that the low--lying bound state energies must remain
constant when the gauge transformation profile $\xi(\rho)$ is
modified. We have confirmed this behavior for simple scale and width
changes in $\xi(\rho)$.

\subsection{Populating the Bound States}

Having determined the set of bound state energies in the
two--dimensional problem, we now integrate these results into the full
three--dimensional calculation. Let $0\le\epsilon_i < m$ represent the
energy of one of these two--dimensional bound states.  In the full
three--dimensional problem, we will then have a family of bound states
with energies $\left[\epsilon_i^2+p_z^2\right]^{\sfrac{1}{2}}$.

For a given charge $Q$, each of these families of bound states will be
filled up to a common chemical potential\footnote{In what follows the
chemical potential $\mu$ should not be confused with the redundant scale
introduced in eq.~(\ref{interface:real}).} $\mu(Q)\le m$, to minimize
their contribution to the energy. If the towers of
states built upon two different $\epsilon_i$ had different upper
limits, the energy would be lowered by moving a state from the tower
with the larger limit to that with the lower one, without changing the
charge. Since states with
$[\epsilon_i^2+p_z^2]^{\sfrac{1}{2}}<\mu$ are filled while states with
$[\epsilon_i^2+p_z^2]^{\sfrac{1}{2}}>\mu$ remain empty, we have a
Fermi momentum $P_i(\mu)= [\mu^2-\epsilon_i^2]^{\sfrac{1}{2}}$ for each bound
state. By the Pauli exclusion principle we can occupy each
state only once, and so we find the charge density per unit length of
the string
\begin{equation}
Q(\mu)=\frac{1}{\pi}\sum_{\epsilon_i\le\mu} P_i(\mu)\,,
\label{eq:charge}
\end{equation}
where the sum runs over all bound states available for a given chemical
potential,\footnote{Ambiguities in this relation
due to different boundary conditions at the end of the string show up
at subleading order in $1/L$, where $L$ is the length of the string,
and can thus be safely ignored.}  $\epsilon_i < \mu$.
Of course, this sum involves
different partial waves, so we have
to include the corresponding degeneracy factors.

Eq.~(\ref{eq:charge}) can be inverted to give $\mu=\mu(Q)$. In  numerical
computations we prescribe the left--hand--side of eq.~(\ref{eq:charge})
and increase $\mu$ from ${\rm min}\{|\epsilon_i|\}$ until the
right--hand--side matches. From this value $\mu = \mu(Q)$,
the binding energy per unit length
\begin{eqnarray}
E_{\rm b}(Q) &=& \frac{1}{\pi}\sum_{\epsilon_i\le\mu}
\int_0^{\mbox{\footnotesize $P_i(\mu)$}}
\hspace{-0.2cm}dp_z \left[\sqrt{\epsilon_i^2+p_z^2}-m\right] \cr
&=& \frac{1}{2\pi} \sum_{\epsilon_i\le\mu} \left[
P_i(\mu) (\mu-2m) + \epsilon_i^2 \ln \frac{P_i(\mu)+ \mu}{\epsilon_i}\right]
\label{eq:ebind}
\end{eqnarray}
can be computed as a function of the prescribed charge. In this manner
the total energy becomes a function of the charge density of the string.
In our search for a stable string, then, we specify the charge $Q$,
and, among background configurations with sufficient binding to
accommodate this charge, we vary the {\it ansatz} parameters to minimize 
the total energy to see if we find a bound configuration.

\section{Numerical Results}
\label{sec:6}

In this section we present the numerical results combining all the contributions 
to the string energy in our variational {\it ansatz}. We measure the variational 
parameters $w_H$ and $w_G$ in inverse fermion masses. The dimensionless vacuum 
polarization energy per unit length $E_q/m^2$ then does not explicitly depend on 
the coupling constants $f$ and $g$ in the $\overline{\rm MS}$ scheme, and depends 
only weakly on these constants in the physical on--shell scheme through 
the logarithmic dependence introduced by the renormalization conditions. This 
property simplifies the numerical analysis because then their variation solely 
affects the classical and counterterm energies, both of which are local functionals 
of the profile functions and hence easy to compute.

We have already presented first results for the vacuum polarization energy,
eq.~(\ref{eq:evac}) in ref.~\cite{Weigel:2010pf}. In particular, we have 
verified our results numerically by checking that they are independent of the shape of
the gauge function $\xi(\rho)$. This result is a consequence of gauge invariance,
but it is nontrivial because the individual Born terms and Feynman diagrams
are not explicitly gauge invariant ---  only the  combination of all of them is.
As a result, this invariance verifies the equivalence between the Feynman diagram
contribution and the Born subtractions (including the fake boson part) in
eq.~(\ref{eq:evac}), which is central to the application of spectral methods
in quantum field theory~\cite{Graham:2009zz}.

The computation of $E_{\delta}$ is numerically most costly. The main
reason is that we have to go to very high angular momenta in the
sum in eq.~(\ref{finalJost}). Typical values are $\ell_{\rm max}=500,\ldots,800$
depending on the width of the background field. To capture the behavior of the
integrand in eq.~(\ref{MASTER2}) we consider about 40 points in the interval
$0\le\tau\le8$. Since the integrand of $E_{\delta}$ does not oscillate when
computed from imaginary momenta, we can accurately estimate the
contribution from  $\tau>8$ from an inverse power--law behavior.

\begin{figure}[tp]
\centerline{~\hspace{-0.5cm}
\includegraphics[width=7.5cm,height=5.0cm]{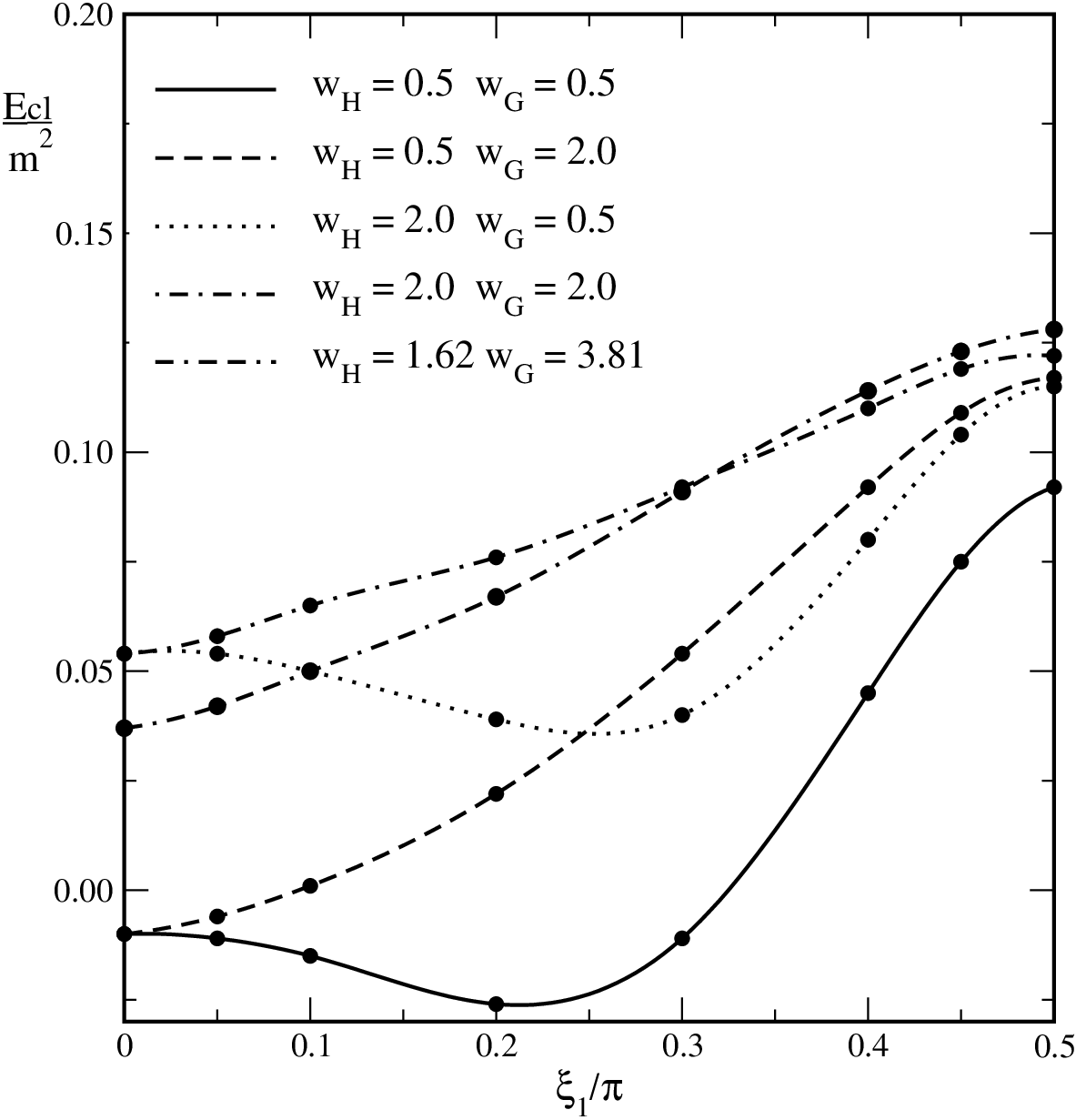}~~
\includegraphics[width=7.5cm,height=5.0cm]{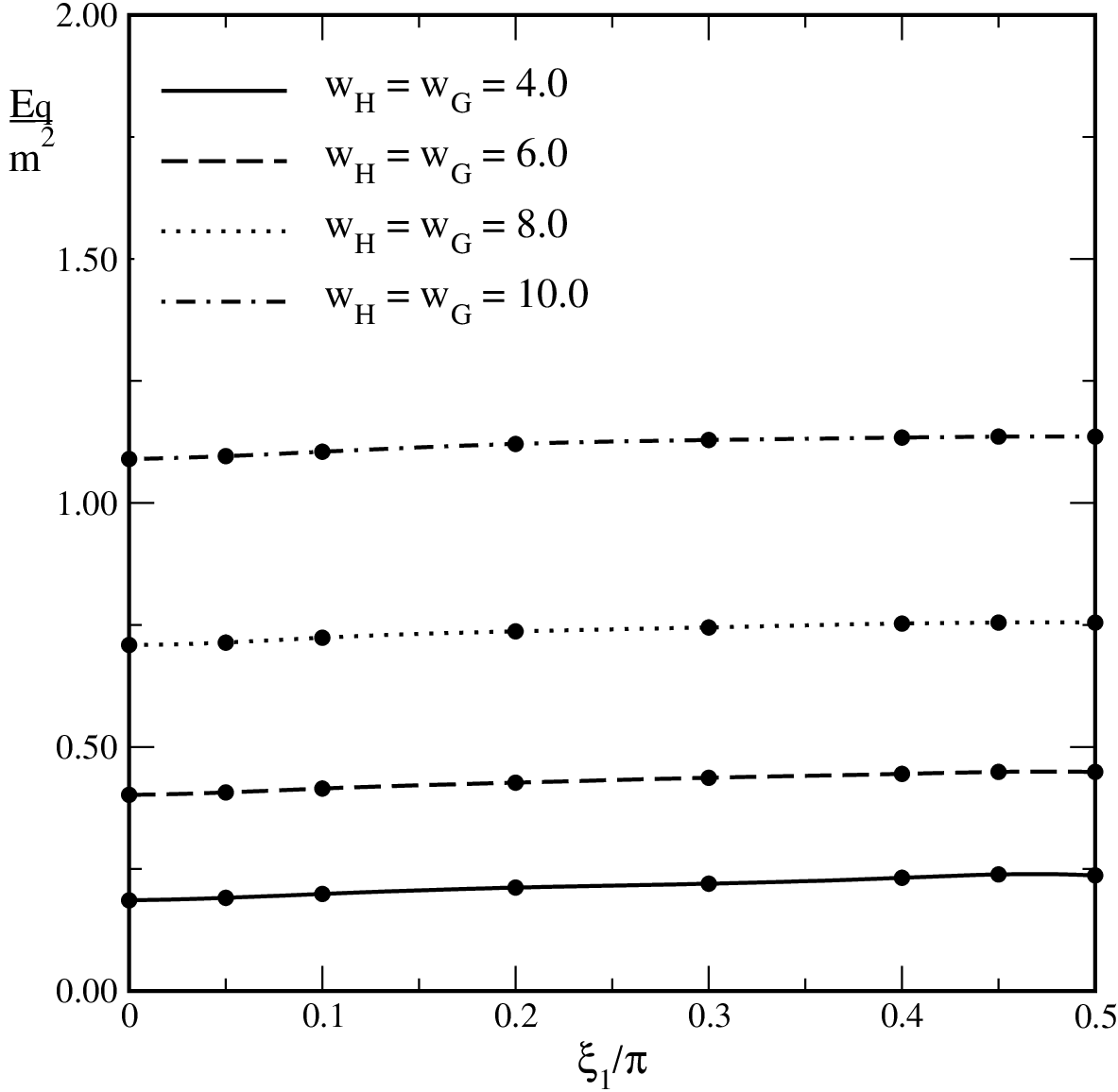}}
\caption{\sf Vacuum polarization energy as function of the angle $\xi_1$
for different values of the width parameters $w_H$ and $w_G$ in the
on--shell renormalization scheme. The physically motivated model
parameters, eq.~(\ref{eq:parameters}), are used. The dots refer to 
actual computations, while the lines stem from a cubic spline. We also 
show the results obtained for the fit to the Nielsen--Olesen profiles,
{\it cf.} figure~\ref{fig:NO}. These results do not include the combinatoric 
color factor $N_C$.}
\label{onshell_fig}
\end{figure}
\begin{figure}[tp]
\centerline{~\hspace{-0.5cm}
\includegraphics[width=7.5cm,height=5.0cm]{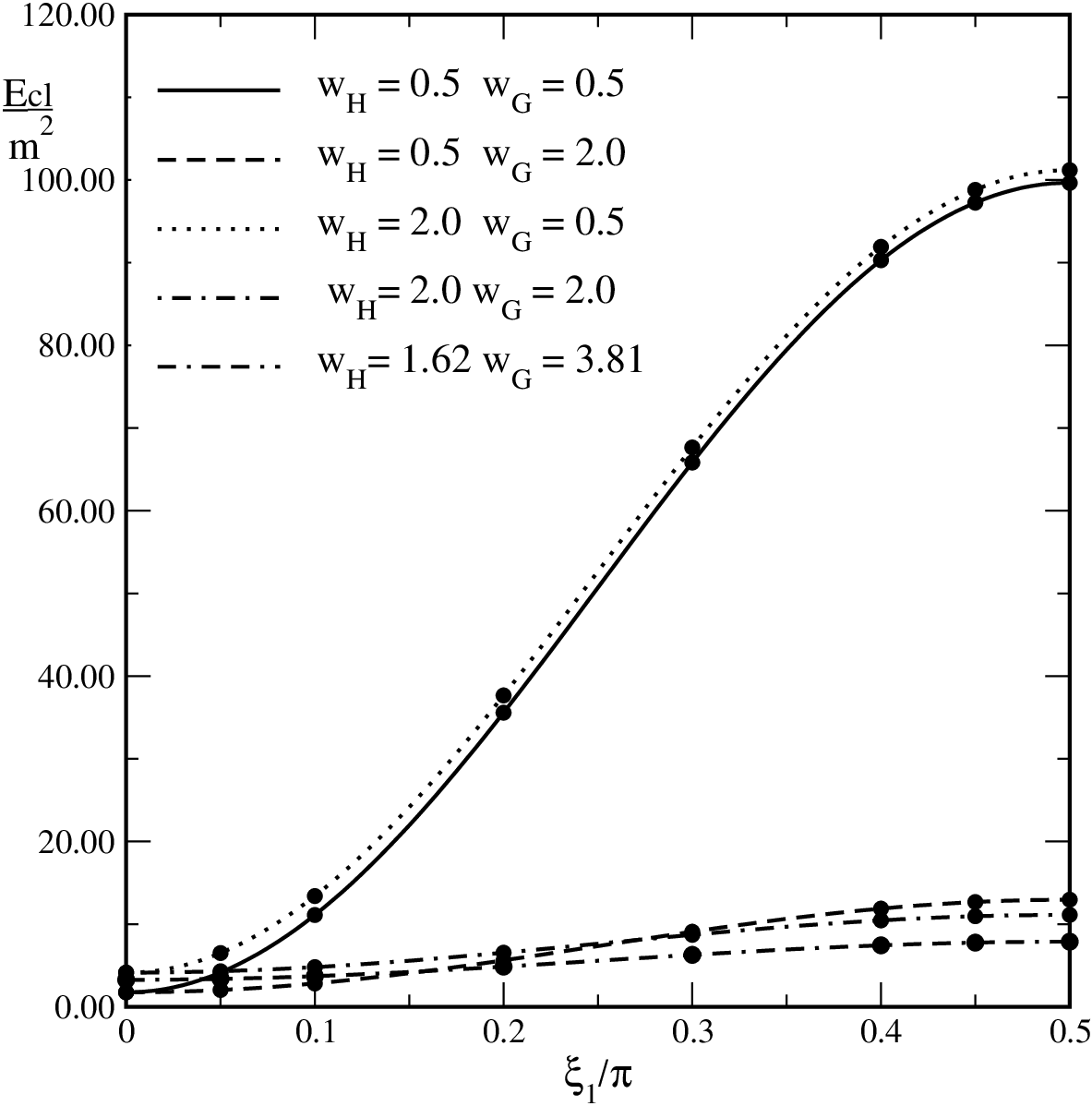}~~
\includegraphics[width=7.5cm,height=5.0cm]{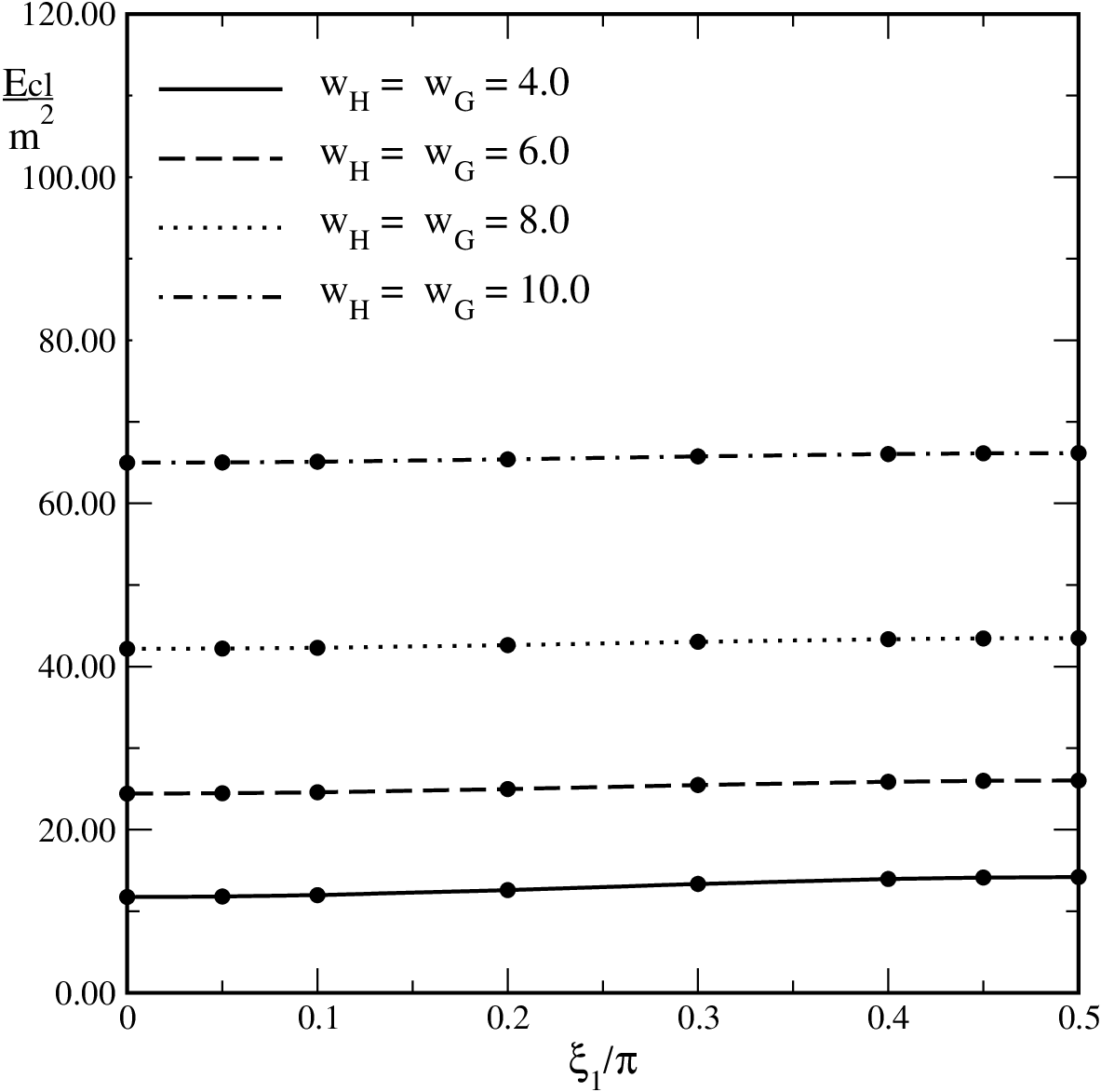}}
\caption{\sf Classical energy for the standard model parameters as function of the
{\it ansatz} parameters $w_H$, $w_G$ and $\xi_1$, including the fit to the 
Nielsen--Olesen profiles. The model parameters are again from
eq.~(\ref{eq:parameters}).}
\label{classical_fig}
\end{figure}
In figure \ref{onshell_fig} we show the result of this
numerical computation.  The wider the background fields, the weaker
the dependence on the angle $\xi_1$  that parameterizes the gauge
boson contribution. Surprisingly, we see that the vacuum polarization per unit
length is quite small. Even for large widths  it does not exceed a fraction of
the fermion mass squared. With the exception of very small widths,
the vacuum polarization turns out to be positive. Hence there is no
indication that the vacuum polarization energy from the fermions can
stabilize cosmic strings since the classical energy is larger by orders of 
magnitude, unless the  coupling constants are $f,g\sim{\mathcal O}(10)$.
For example, see figure~\ref{classical_fig}, which shows the
classical energy for the standard model parameters, which are ${\mathcal O}(1)$.
The derivative terms of the classical energy decrease quadratically with $f$ 
and $g$ while the Higgs potential decreases  like $1/f^4$ for fixed Higgs mass.
As a result, increasing the coupling constants could lead to binding for thin 
strings, but such configurations contain large Fourier components, which 
for $f,g\sim {\cal O}(10)$ reach the vicinity of the Landau ghost pole.
Hence any such binding is obscured by the existence of the Landau ghost, 
{\it cf.} appendix C.4, which arises when including
quantum corrections in a manner that does not reflect asymptotic
freedom. Here it is due to the omission of  quantum corrections from
fluctuating gauge boson fields. The estimate for the Landau ghost
contribution discussed in the appendix suggests that the issue can be safely
ignored for $f,g\lesssim5$.

Gradient expansions~\cite{Kir93} for quantum mechanical expectation
values suggest that the energy gain from populating bound states can
be estimated from a spatial integral over some fractional power of the
potential in the wave--equation. Scaling arguments show that the
energy from the populated bound states increases quadratically with
the width parameters $w_H$ and/or $w_G$, regardless of the specific
power in the expansion\footnote{More precisely, the energy gain
involves both the summed energy eigenvalues and the charge. Both can
be expressed by such integrals with different powers, though.}. Since
also the dominating classical energy increases quadratically with
$w_H$ (from the Higgs potential), populating the bound states might
balance the large classical energy already at small coupling constants
and moderate widths, because the Higgs
potential scales like $1/f^4$ when the Higgs mass is fixed while the fermion
contribution is not sensitive to any change in $f$
when we scale our definitions of physical quantities
with the fermion mass $m=vf$. Hence our
strategy to construct a stable string type configuration is to balance
the classical boson energy with the fermion quantum correction by
considering wide strings and increasing the Yukawa coupling.
Simultaneously we must keep fixed the charge associated with
populating the fermion bound states.

\begin{figure}[tp]
\centerline{~\hspace{-0.5cm}
\includegraphics[width=10.0cm,height=5.0cm]{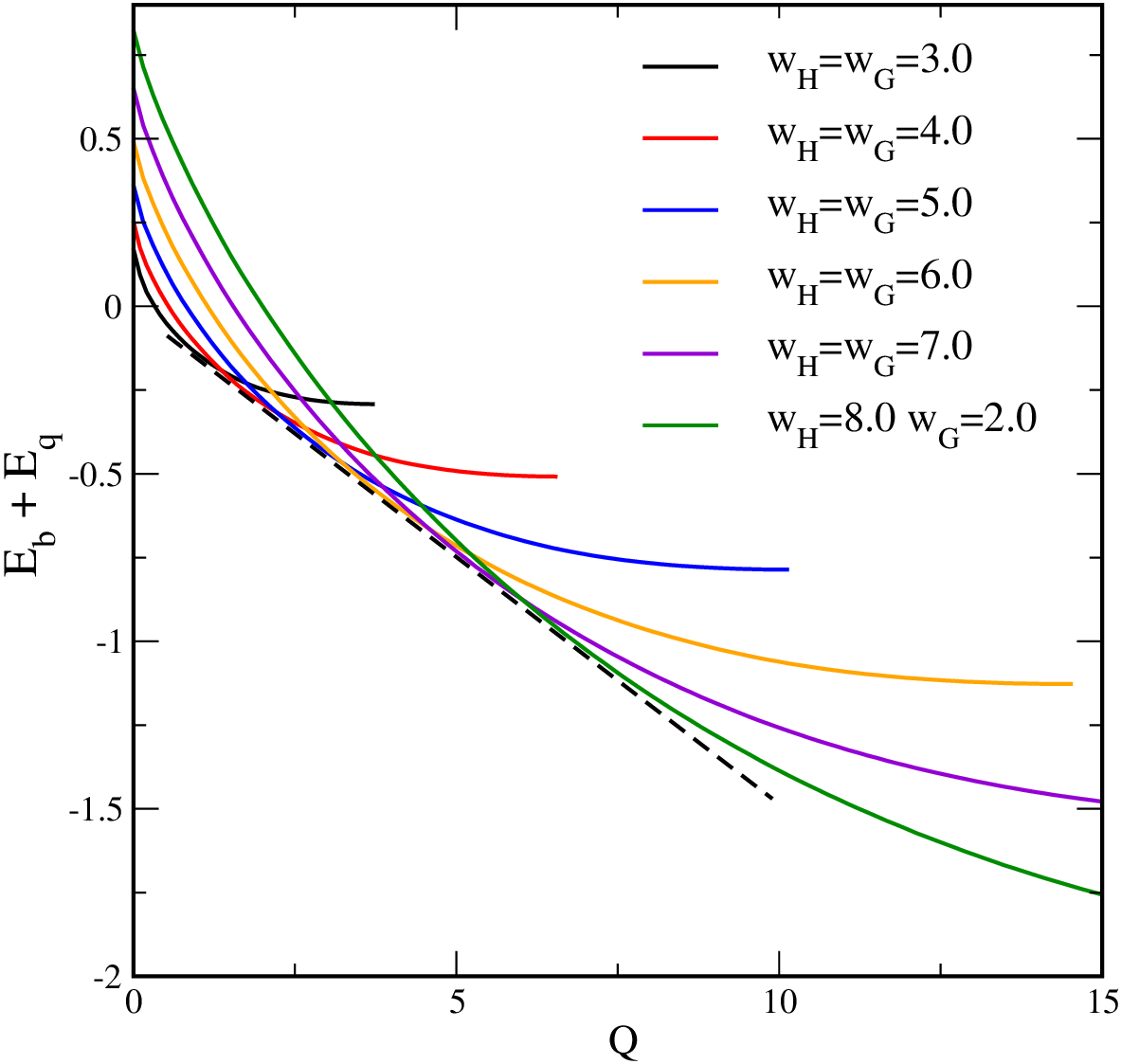}}
\caption{\sf (Color online) Total bound state and vacuum energy per unit
length as a function of charge density per unit length, in units of
the fermion mass, for $\xi_1=0.4\pi$. The dotted line
indicates the minimal fermionic contribution to the energy.}
\label{fig:ebline}
\end{figure}

Before we will consider the total energy we would like to discuss the
fermion part, $E_{\rm q}+E_{\rm b}$. In figure~\ref{fig:ebline}
we show the total fermion energy of eq.~(\ref{energy:quantum}), as a
function of the charge density per unit length of
the string.  As described in the previous section, we can
compute the binding energy as a function of the charge for a given
background configuration. After  adding the vacuum polarization energy
computed for that background, we get the
parabolic curves in figure~\ref{fig:ebline}. These lines terminate at
the point where all available bound states are populated.
We then search for the configuration that minimizes the energy.
For small charges, we obtain thin strings, while larger charges lead to wide
strings, as shown in figure~\ref{fig:ebline}.  Surprisingly,
the resulting envelope that describes the minimal fermion energy as a
function of the charge density is a straight line with (approximately)
vanishing $y$--intercept. This straight line stems from a delicate
balance between the vacuum  polarization and binding energies.
Because this extrapolation yields a vanishing $y$--intercept, we
deduce that very narrow strings have vanishing vacuum polarization
energy. This interpolation overcomes the Landau ghost problem of the
direct calculation.

\begin{figure}[tp]
~\vspace{0.5cm}\\
\centerline{~\hspace{-0.5cm}
\includegraphics[width=10.0cm,height=5.0cm]{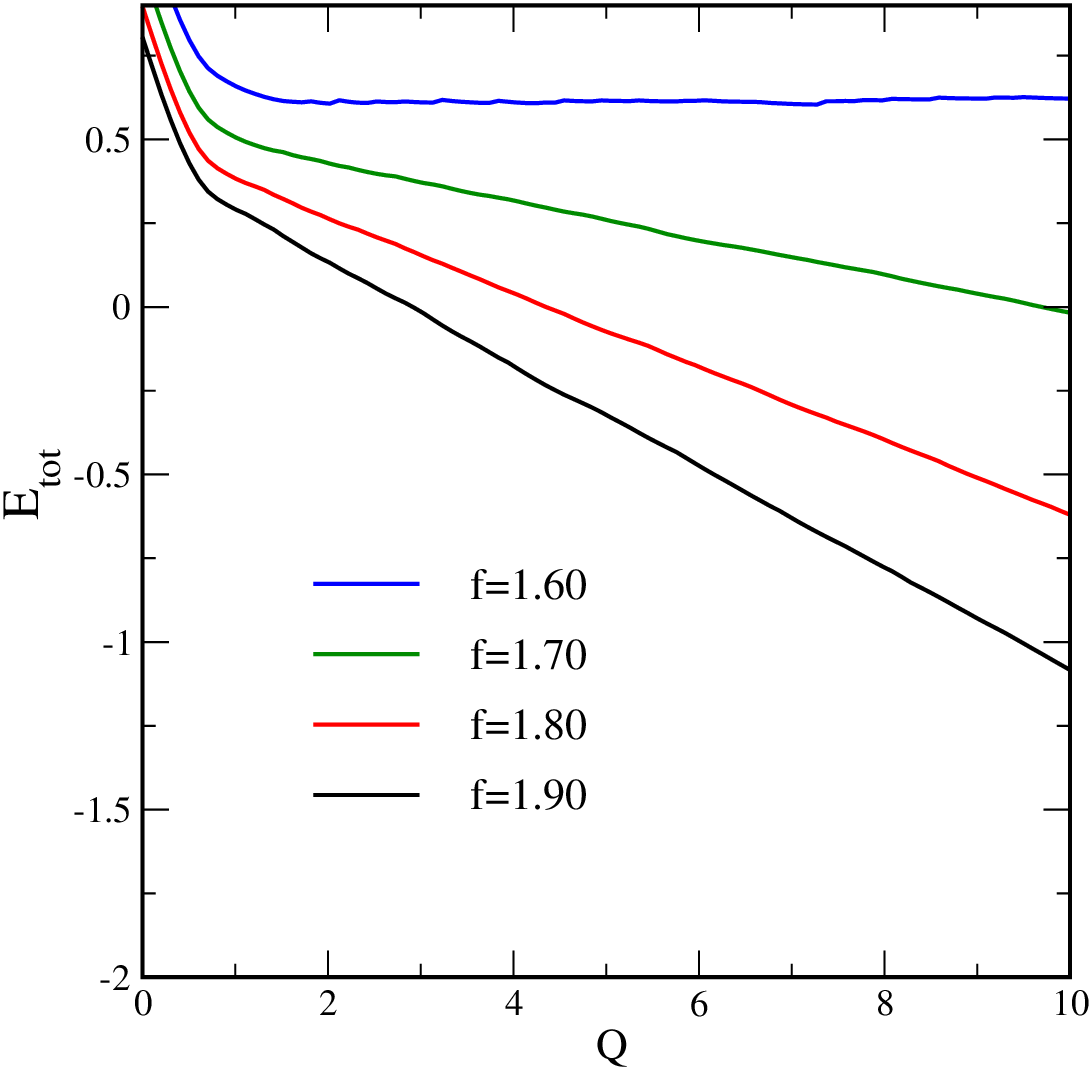}}
\caption{\sf (Color online) Total energy as a function of the charge for various
values of the Yukawa coupling constant. The color degeneracy $N_C=3$ is included.}
\label{fig:etot}
\end{figure}

From several hundred configurations for which we have computed both
the vacuum polarization energy and constructed the bound states, we 
identify the one that minimizes the total binding energy 
\begin{equation}
E_{\rm tot}=E_{\rm cl}+E_{\rm q}+E_{\rm b}
\label{etot}
\end{equation}
for a prescribed charge. The corresponding result for the minimal binding
energy is displayed in figure~\ref{fig:etot}.
We can see that the optimal binding grows linearly with $Q$.
The steep slope at very  small charges is an artifact of restricting our
{\it ansatz} to configurations with $w_H, w_G\ge2$ to avoid unphysical effects
from the Landau ghost.

As mentioned above, we increase the Yukawa coupling from its
top--quark motivated  value $f=0.99$, while all other model
parameters are taken from eq.~(\ref{eq:parameters}).
Increasing the charge also increases the width of the optimal
string. For $f \approx 1.6$ the classical and quantum
contributions balance and the total energy is essentially
independent of the width. Increasing the coupling further yields a
negative energy (in comparison to equally many free fermions) and
stable configurations exist.  Not surprisingly, the minimal charge for
which there are stable configurations decreases quickly as $f$
increases. For $f=1.7$ it is  $Q_{\rm min}\sim 10m=17v$, while for
$f=1.9$ stable configurations exist already at $Q_{\rm min}\sim3m=5.7v$.

\begin{figure}[tp]
~\vspace{0.5cm}\\
\centerline{~\hspace{-0.5cm}
\includegraphics[width=10.0cm,height=5.0cm]{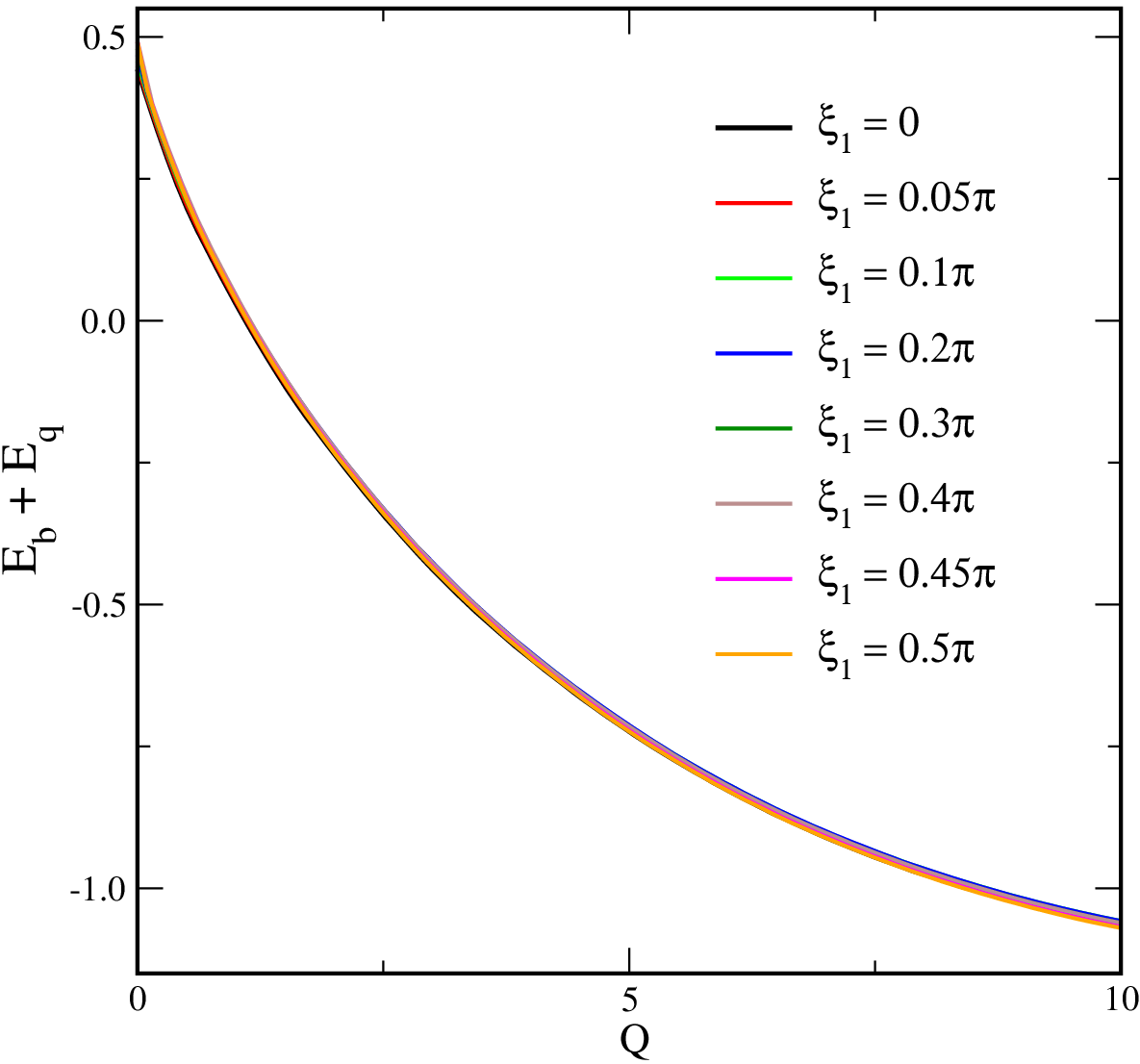}}
\caption{\sf (Color online) Fermionic contribution to the string binding
energy per unit length as a function of charge density per unit
length, in units of the fermion mass, for a variety of values of $\xi_1$
with $w_H=6.0$ and $w_G=6.0$.}
\label{fig:xi1}
\end{figure}
We next discuss the structure of the stabilizing configuration.
We find that the fermionic part of the binding energy
is insensitive to the angle $\xi_1$, as shown in figure~\ref{fig:xi1}.
As a result, the dependence of the total binding energy on $\xi_1$ stems
entirely from the classical part, which is clearly minimized
for $\xi_1\sim 0$, since in that case the the gauge fields vanish and we
have only a charged Higgs field, with only the non--diagonal elements
in eq.~(\ref{string_higgs}) differing from zero.

\begin{figure}[tp]
~\vspace{0.5cm}\\
\centerline{~\hspace{-0.5cm}
\includegraphics[width=10.0cm,height=5.0cm]{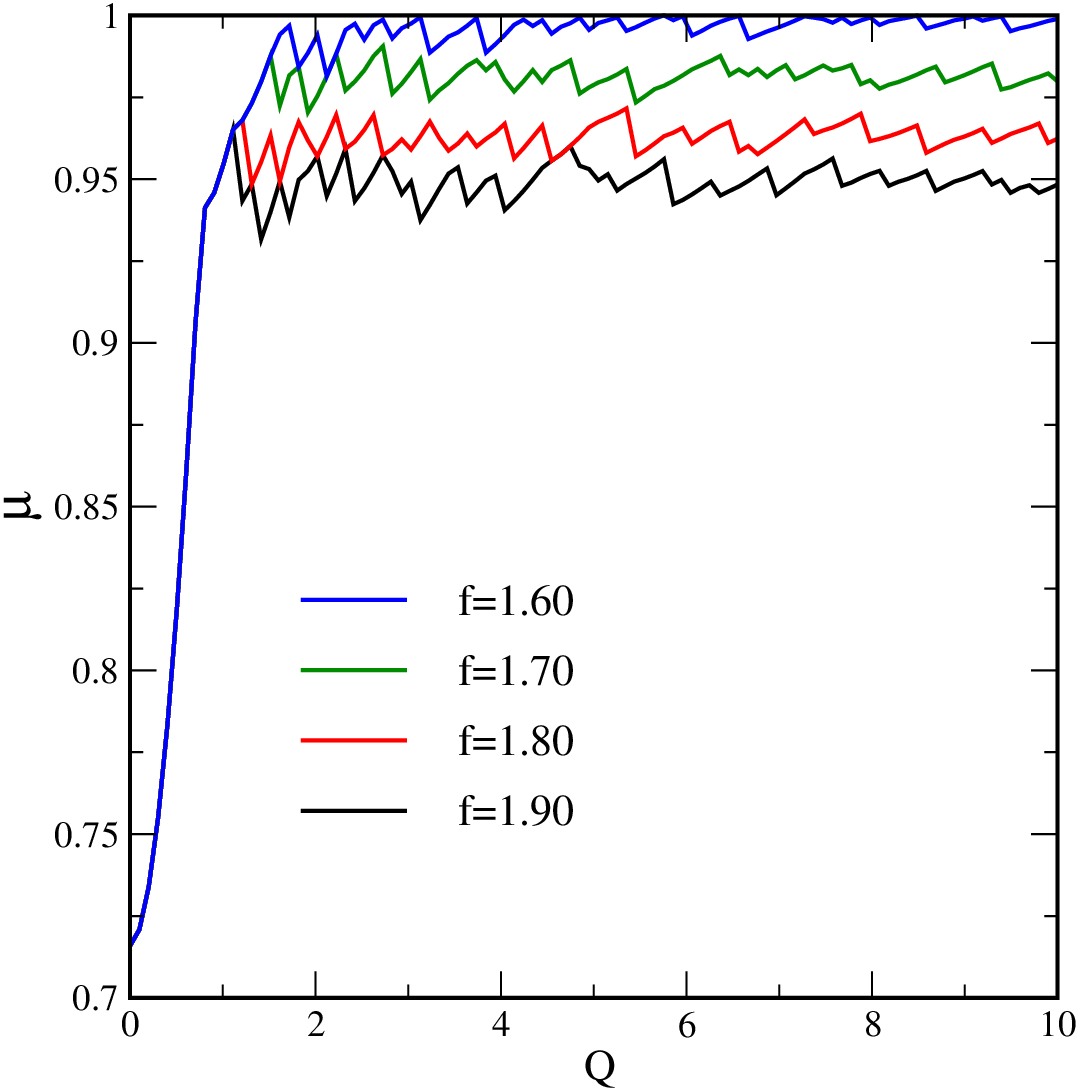}}
\caption{\sf (Color online) The chemical potential that minimizes the
binding energy for a prescribed charge in units of the fermion mass.}
\label{fig:mu}
\end{figure}

In figure \ref{fig:mu} we display the chemical potential that minimizes
the binding energy for a prescribed charge. Its construction is
discussed in section~\ref{sec:5}.
The cusps arise because our sample configurations are not continuous in the
variational parameters. As we increase the charge, the minimizing
configuration jumps among these possibilities.

The strong deviation from $\mu=m$ at low charges (where the various graphs
overlap) is again an artifact of not considering very narrow string
configurations. We see that at the limit of binding ($f=1.6$) almost
all bound states are populated. As the binding increases, the chemical
potential decreases, leaving the states below threshold un--occupied.

\begin{table}
\centerline{
\begin{tabular}{c|cccc}
$f\Bigg\backslash Q $ & 2 & 4 & 6 & 8 \cr
\hline
1.6~ & 2.3 & 3.2 & 4.0 & 4.5\cr
1.7~ & 2.5 & 3.5 & 4.3 & 5.0\cr
1.8~ & 2.7 & 3.8 & 4.6 & 5.5\cr
1.9~ & 2.8 & 4.2 & 5.1 & 5.8
\end{tabular}}
\caption{\label{tab:width}\sf The width, $w_H$, of the minimizing Higgs
profile at a prescribed charge, $Q$ for different values of the Yukawa
coupling constant $f$.}
\end{table}
We have seen above that binding increases with the Yukawa coupling.
Table \ref{tab:width} indicates that at the same time the profile functions
get wider, while the critical charge at
which binding sets in ({\it i.e.} $E_{\rm tot}(Q_{\rm min})=0$)
decreases with the Yukawa  coupling.  As a result,
the width of the critical profile actually decreases. We  find these
widths to be $5.5/m=3.2/v$, $4.0/m=2.2/v$ and $3.5/m=1.8/v$ for
$f=1.7$, $1.8$ and $1.9$, respectively, and the
typical extension of a bound charged string is about $0.003{\rm fm}$.

Finally let us estimate the total mass of the bound string. In the regime
where it is only slightly bound, we have $E_{\rm tot}\lesssim Qm$. Typically
we observe binding for $Q\approx 5m$. Hence a reasonable estimate for the
mass of the string is $M\approx 5m^2L$. Taking $m=300{\rm GeV}$ and the length
of the string to be the radius of the sun,
$L=R_\odot\approx7\times10^8{\rm m}$ we find
$M\approx2.3\times10^9{\rm kg}=10^{-20}M_\odot$, {\it i.e.} only a
very tiny fraction of the mass of the sun. 
On the microscopic scale,
a string as short as the Compton wave--length of the heavy fermion
would carry about 30 bound fermions and have an energy of slightly
less than $9{\rm TeV}$.

\section{Conclusions}
\label{sec:7}

We have extended our previous spectral approach to find the leading
quantum corrections to the energy of a cosmic string in a slightly simplified
version of the electroweak theory. In the limit of many
internal degrees of freedom,  $N_C \to\infty$, these leading
corrections come from fermions coupling to the string background.
In this scenario $N_C$ merely appears a combinatoric
factor, which is justified by the asymptotic freedom of QCD.
We have shown how to compute the distortion of the Dirac spectrum
in the string background, and how to extract the full non--perturbative
renormalized vacuum polarization using perturbative counterterms and
conventional renormalization schemes. Substantial refinements of our
previous techniques were necessary to make this calculation feasible,
and we have presented a complete account including technical
details in the appendices. Though we have focused on
the computational method underlying previously published results,
we have also discussed some novel results concerning the
structure of the stable configuration.

The basic idea of the quantum stabilization of cosmic strings is that the
appearance of (near) zero modes in the distorted Dirac spectrum could
help to produce negative contributions to the energy that overcomes 
the classical energy necessary to form
the string. We have shown, however, that the contribution from the distortion
of the remaining parts of the spectrum, {\it i.e.} the scattering states,
neutralizes the binding effect of the low--lying modes, resulting
in a very small vacuum polarization energy. In particular,
the vacuum is stable against spontaneous formation of weak strings for
parameters that are physically sensible. The situation is more favorable for
\emph{charged} strings with explicitly occupied bound states, since such
configuration need only be lighter than the same number of free fermions,
to be stable on time scales over
which we may neglect fermion number nonconservation.
This approximation is valid if the inter--generation quark mixing is tiny 
(we have assumed it to be zero), and if the quark masses in the heavy 
fermion doublet are nearly degenerate (we have assumed exact degeneracy).

For otherwise realistic parameters, we have shown that this binding mechanism
sets in at surprisingly low fermion masses of around $300\,\mathrm{GeV}$.
This corresponds to values of the Yukawa coupling that are
still small enough for our calculations based on the Standard Model
to be reliable.  A stable charged string can thus be
formed when enough charge density of a heavy fermion
doublet with about twice the top quark mass is available.

If taken at face value, our findings suggest that a weakly coupled fourth 
generation of heavy quarks would make its footprint through the electroweak 
string phenomenology mentioned in the introduction --- or conversely, that 
the non--observation of electroweak strings would put severe bounds on the 
masses of possible heavy quarks.  However, such conclusions must be qualified 
by a number of simplifications that were necessary to make the calculation
feasible.  Most notably, the restriction to fermionic quantum fluctuations, 
although justified by the large $N_C$--argument, leads to a quantum theory 
that is not asymptotically free, and in turn to the Landau pole problem at 
small string widths. We have presented a crude way of estimating this
contribution in order to ensure that our findings are not affected by
it. This treatment should obviously be improved by a full quantum
calculation of the bosonic contribution to the vacuum polarization
energy. Recent studies~\cite{Baacke:2008sq} in models
related to ours indicate that the bosonic contribution can give interesting
and non--trivial effects. We are currently investigating such an extension
of our model.

Other shortcomings of our model are the mass degeneracy of the fermion 
doublet, the lack of inter--generation quark couplings and, in particular,
the decoupling of $U(1)$ hypercharge. While the stable configuration that we have
constructed can be embedded in a full $SU(2) \times U(1)$ model with
multiple generations (since additional degrees of freedom would only 
serve to lower the energy in our variational approach), it is
unclear if the new couplings provide new decay channels, in particular
when bosonic fluctuations are taken into account. We also plan to
investigate such a scenario.

The charges along the string may carry 
currents~\cite{Davis:1988jq,Davis:1995kk,Davis:1996xs}, which in turn
can have interesting consequences for baryogenesis and
cosmology~\cite{Davis:1999ec,Lilley:2010av}. This situation 
is similar to the Witten model \cite{Witten:1984eb,Jackiw:1981ee}
and its generalizations, where the currents are
induced by the coupling of extra scalar fields to the vortex.
In this scenario, the Brownian network of vortices produced in an earlier
(GUT-scale) phase transition contracts as the universe cools down.
This process could eventually be stopped by the currents becoming
superconducting, with the irregular vortex shapes being smoothed out by the
surrounding thermal background to form circular rings. The
final evolution stage would then be a universe filled with microscopic
superconducting, charged vortex loops. Such a 
\emph{vorton}~\cite{Davis:1988ij,Brandenberger:1996zp} universe has recently
attracted  much attention because it provides a viable candidate for dark matter with
rather accurately computable properties that put stringent restrictions on
cosmological models. It would be very interesting to study such a possibility 
in the electroweak standard model, with currents produced directly from fermions 
(as they are in our calculation), rather than from extra 
scalar fields. Although our present investigation does not directly address
this question, it seems conceivable that a stable vorton could be created without 
requiring exceedingly large couplings or unrealistic masses. 
Combining this scenario to our picture could be another avenue for 
future research.

Finally, it would of course also be interesting to study the Brownian network
of strings as it is formed in the phase transition if enough fermion charge is
available. Due to their complexity, such configurations must presumably be
studied in an effective (lattice) model. The necessary string interactions could
potentially be addressed through further extensions of the spectral method.

\acknowledgments
N.~G.\ is supported in part by the NSF through grant PHY08-55426.

\appendix

\section{Eigenvalue Problem}
\label{appA}

To find the bound state spectrum, states with energy eigenvalues $|\epsilon|<m$, 
we first diagonalize the Hamiltonian matrix in the absence of a background 
potential, and then use these free eigenstates (with the proper boundary 
conditions built in) as a basis in which to compute the matrix elements 
of the background potential induced by the string. The diagonalization of 
this full Hamiltonian matrix in turn yields the fully interacting bound 
state spectrum.  Note that in this procedure all states appear as
``bound'' states, since the volume of the coordinate space is finite.
In general the energy eigenvalues of such ``bound'' states depend on
the volume.  However, the \emph{true} bound states with $|\epsilon|<m$ do not
show finite size effects if the volume is chosen large enough because 
their wave--functions are located in a small sub--volume.

The single particle Dirac Hamiltonian couples spin ($S$) and
weak isospin ($I$) degrees of freedom. We can combine
these degrees of freedom by introducing the \emph{grand spin} states 
given in eq.~(\ref{eq:GSstates}). The matrix elements of the (two--component) 
operators entering the Hamiltonians in eqs.~(\ref{eqDirac}) and (\ref{hint1})
are listed in tables \ref{tabSR} to \ref{tabIP}. In all of these tables,
we use the abbreviation $s_x={\rm sin}(x)$ and $c_x={\rm cos}(x)$,
where the arguments of these trigonometric functions appear as subscripts.


\begin{table}
\begin{minipage}[t]{0.4\linewidth}
\centerline{
\begin{tabular}{c|cccc}
states & $|++\rangle$ & $|+-\rangle$ & $|-+\rangle$ & $|--\rangle$ \cr
\hline
$\langle ++|$ & 0 & 0 & -1& 0 \cr
$\langle +-|$ & 0 & 0 & 0 & -1\cr
$\langle -+|$ & 1 & 0 & 0 & 0 \cr
$\langle --|$ & 0 & 1 & 0 & 0 \cr
\end{tabular}}
\caption{\label{tabSR} \sf Matrix elements of
$i\Vek{\sigma}\cdot\widehat{\mbox{\boldmath$\rho$}}$. The table should be read as
$\langle\,-+|\,i\Vek{\sigma}\cdot\widehat{\mbox{\boldmath$\rho$}}\,|++\,\rangle=1$, for instance.
The matrix is hermitian in combination with the derivative $\partial_\rho$.}
\end{minipage}
\hspace{2.0cm}
\begin{minipage}[t]{0.4\linewidth}
\centerline{
\begin{tabular}{c|cccc}
states & $|++\rangle$ & $|+-\rangle$ & $|-+\rangle$ & $|--\rangle$ \cr
\hline
$\langle ++|$ &     0       &     0   & $-(\ell+n+1)$ &      0      \cr
$\langle +-|$ &     0       &     0   &        0      & $-(\ell+1)$ \cr
$\langle -+|$ & $-(\ell+n)$ &     0   &        0      &      0      \cr
$\langle --|$ &     0       & $-\ell$ &        0      &      0      \cr
\end{tabular}}
\caption{\label{tabSPhi} \sf Matrix elements of
$i\,(\Vek{\sigma}\cdot\widehat{\mbox{\boldmath$\varphi$}})\,\partial_{\varphi}$.}
\end{minipage}
\end{table}

\begin{table}
\begin{minipage}[t]{0.4\linewidth}
\centerline{
\begin{tabular}{c|cccc}
states & $|++\rangle$ & $|+-\rangle$ & $|-+\rangle$ & $|--\rangle$ \cr
\hline
$\langle ++|$ & 0 & 0 & 0 & $1$  \cr
$\langle +-|$ & 0 & 0 & $-1$ & 0  \cr
$\langle -+|$ & 0 & $-1$ & 0 & 0 \cr
$\langle --|$ & $1$ & 0 & 0 & 0 \cr
\end{tabular}}
\caption{\label{tabrhoIP}\sf Matrix elements of
$(\Vek{\sigma}\cdot\widehat{\mbox{\boldmath$\rho$}})\,I_P$.}
\end{minipage}
\hspace{2cm}
\begin{minipage}[t]{0.4\linewidth}
\centerline{
\begin{tabular}{c|cccc}
states & $|++\rangle$ & $|+-\rangle$ & $|-+\rangle$ & $|--\rangle$ \cr
\hline
$\langle ++|$ &  0   &  0   & $-\sd$ & $-\cd$ \cr
$\langle +-|$ &  0   &  0   & $-\cd$ & $\sd$ \cr
$\langle -+|$ & $-\sd$ & $-\cd$ &  0   &  0   \cr
$\langle --|$ & $-\cd$ & $ \sd$ &  0   &  0   \cr
\end{tabular}}
\caption{\label{tabIG_D}\sf Matrix elements of
$(\Vek{\sigma}\cdot\widehat{\mbox{\boldmath$\varphi$}})\,I_G(\Delta)$. 
The subscript denotes the argument of the trigonometric functions.}
\end{minipage}
\end{table}

\begin{table}
\begin{minipage}[t]{0.45\linewidth}
\centerline{
\begin{tabular}{c|cccc}
states & $|++\rangle$ & $|+-\rangle$ & $|-+\rangle$ & $|--\rangle$ \cr
\hline
$\langle ++|$ &  0   &  0   & $\sx$  & $-\cx$ \cr
$\langle +-|$ &  0   &  0   & $-\cx$ & $-\sx$ \cr
$\langle -+|$ & $ \sx$ & $-\cx$ &  0   &  0   \cr
$\langle --|$ & $-\cx$ & $-\sx$ &  0   &  0   \cr
\end{tabular}}
\caption{\label{tabIG_x}\sf Matrix elements of
$(\Vek{\sigma}\cdot\widehat{\mbox{\boldmath$\varphi$}})\,I_G(-\xi)$.
The subscript denotes the argument of the trigonometric functions.}
\end{minipage}
\hspace{2cm}
\begin{minipage}[t]{0.4\linewidth}
\centerline{
\begin{tabular}{c|cccc}
states & $|++\rangle$ & $|+-\rangle$ & $|-+\rangle$ & $|--\rangle$ \cr
\hline
$\langle ++|$ & 0 & $1$  &  0 &  0  \cr
$\langle +-|$ & $-1$ & 0 &  0 &  0  \cr
$\langle -+|$ &  0 & 0   &  0 & $1$ \cr
$\langle --|$ &  0 & 0   & $-1$ & 0 \cr
\end{tabular}}
\caption{\label{tabIP}\sf Matrix elements of $iI_P$. The extra factor
$i$ leads to anti--hermitian matrix elements, compensating the same
property of its spinor coefficient $\beta\gamma_5$.}
\end{minipage}
\end{table}

\noindent
Next, radial functions are introduced via the four--component spinors,
{\it cf.} eq.~(\ref{eq:GSspinors}),
$$
~\hspace{2cm}
\begin{array}{r@{\,\,\,=\,\,\,}l@{\qquad\longrightarrow\qquad}l}
\displaystyle \langle \rho |\epsilon\,\ell++\rangle
&
\displaystyle { J_{\ell+n}(k\rho)|\ell + +\rangle \choose
\kappa J_{\ell+n+1}(k\rho)|\ell - +\rangle }
&
\displaystyle { f_1(\rho)|\ell + +\rangle \choose
g_1(\rho)|\ell - +\rangle }
\\[6mm]
\displaystyle \langle \rho |\epsilon\,\ell +-\rangle &
\displaystyle { J_{\ell}(k\rho)|\ell + -\rangle \choose
\kappa J_{\ell+1}(k\rho)|\ell - -\rangle }
&
\displaystyle { f_2(\rho)|\ell + -\rangle \choose
g_2(\rho)|\ell - -\rangle }
\\[6mm]
\displaystyle \langle \rho |\epsilon\,\ell-+\rangle &
\displaystyle  {J_{\ell+n+1}(k\rho)|\ell - +\rangle \choose
\kappa J_{\ell+n}(k\rho)|\ell + +\rangle }
&
\displaystyle { f_3(\rho)|\ell - +\rangle \choose
g_3(\rho)|\ell + +\rangle }
\\[6mm]
\displaystyle \langle \rho |\epsilon\,\ell--\rangle
&
\displaystyle {J_{\ell+1}(k\rho)|\ell - -\rangle \choose
\kappa J_{\ell}(k\rho)|\ell + -\rangle }
&
\displaystyle { f_4(\rho)|\ell - -\rangle \choose
g_4(\rho)|\ell + -\rangle }\,.
\hspace{2.2cm} \mbox{\refstepcounter{equation}\label{eq:GSspinors_app}(\theequation)}
\end{array}
$$
We note that $\kappa={\rm sgn}(\epsilon)\,\sqrt{\frac{\epsilon-m}{\epsilon+m}}$ 
is well defined for either sign of the energy eigenvalue since $|\epsilon|>m$.
Using the dispersion relation for real momenta $\epsilon^2=k^2+m^2$, we
may also write $\kappa=\frac{k}{\epsilon+m}=\frac{\epsilon-m}{k}$. These two 
expressions are odd in $k$ and thus suitable for analytic continuation $k\to it$.
The spinors to the left of the arrows in eq.~(\ref{eq:GSspinors_app})
involve ordinary Bessel functions which solve the free Dirac
equation. They will
be used to construct the basis states for the Hamiltonian matrix.

In the free case, the four spinors in eq.~(\ref{eq:GSspinors_app}) each solve
the Dirac equation individually, {\it i.e.} they do not couple.
Once the background potential from the string is
included, however, the radial functions $f_i$ and $g_i$
become distorted and mix under the dynamics.

We now construct a discrete basis built from
solutions of the free Dirac equation. To this end we must first impose the
boundary condition that \emph{no flux} runs from the center of the string
through a circle at a large distance $R$. Since the flux is bilinear in the
spinors with all products involving both an upper and a lower component, the
no--flux boundary condition is equivalent to the requirement that
either component vanishes. For the string winding $n=1$, this amounts
to the simple statement
\begin{equation}
J_{\ell+1}(k_r^{(\ell)}R)=0\,.
\label{discret1}
\end{equation}
This conditions selects discrete momenta $k_r^{(\ell)}$ in each
angular momentum channel $\ell$, where $r=1,2,\ldots$ enumerates the momenta
and thus the free basis states. We note in passing that a string winding
$n\ge2$ would require two separate sets of discrete momenta.

The normalization of the spinors can be worked out using the Bessel function
identity
\begin{equation}
\int_0^1 t dt J_\nu(\lambda_r^{(\nu)} t)
J_\nu(\lambda_s^{(\nu)} t) =
\frac{1}{2} \left[J_\nu^\prime(\lambda_r^{(\nu)} t)\right]^2 \delta_{rs}
\label{norm}
\end{equation}
where $\lambda_r^{(\nu)}$ are the roots of the Bessel function $J_\nu$.
Using furthermore the recursion relations for Bessel functions and their
derivatives, we arrive at the following explicit expressions for the
(free) radial functions in equation~(\ref{eq:GSspinors}),
\begin{equation}
\begin{tabular}{lcllcl}
$f_1^{(r)}(\rho)$ & = & $N^{(r)}_f J_{\ell+1}(k_r\rho)$ ~~~~~~ &
$g_1^{(r)}(\rho)$ & = & $N^{(r)}_g J_{\ell+2}(k_r\rho)$ \\[2mm]
$f_2^{(r)}(\rho)$ & = & $N^{(r)}_f J_{\ell}(k_r\rho)$ ~~~~~~ &
$g_2^{(r)}(\rho)$ & = & $N^{(r)}_g J_{\ell+1}(k_r\rho)$ \\[2mm]
$f_3^{(r)}(\rho)$ & = & $N^{(r)}_f J_{\ell+2}(k_r\rho)$ ~~~~~~ &
$g_3^{(r)}(\rho)$ & = & $N^{(r)}_g J_{\ell+1}(k_r\rho)$ \\[2mm]
$f_4^{(r)}(\rho)$ & = & $N^{(r)}_f J_{\ell+1}(k_r\rho)$ ~~~~~~ &
$g_4^{(r)}(\rho)$ & = & $N^{(r)}_g J_{\ell}(k_r\rho)$
\end{tabular}
\label{radfunction}
\end{equation}
where the superscripts on the momenta are omitted. The normalization
factors are given explicitly by
\begin{equation}
N_f^{(r)}=\frac{1}{R}\frac{1}{|J_{\ell+2}(k_rR)|}
\sqrt{\frac{\epsilon_r+m}{\epsilon_r}}\,,
\quad\qquad
N_g^{(r)}=\frac{1}{R}\frac{{\rm sgn}(\epsilon_r)}{|J_{\ell+2}(k_rR)|}
\sqrt{\frac{\epsilon_r-m}{\epsilon_r}}\,.
\label{normrad}
\end{equation}
To limit the number of basis states, we introduce a cutoff
$\Lambda$ and only include momenta with $k_{r} < \Lambda$.
This defines $r_{\rm max}$, the maximal number of discrete
momenta, which depends on $R$ for fixed $\Lambda$.
Due to the energy degeneracy this cutoff truncates the label $r$
on the energy eigenvalues to run from $1\ldots 2 r_{\max}$,
\begin{equation}
\epsilon_r=\begin{cases}
-\sqrt{k_{r_{\rm max}+1-r}^2+m^2} & \quad r=1,\ldots,r_{\rm max} \\[4mm]
\sqrt{k_{r-r_{\rm nmax}}^2+m^2} & \quad r=r_{\rm max}+1,\ldots,2r_{\rm max}\,.
\end{cases}
\label{energystates}
\end{equation}
Putting all the pieces together, we can now present the full Hamiltonian
matrix  {\it i.e.} the operator in eq.~(\ref{eqDirac}) sandwiched
between the spinors constructed above. The equations become simpler if we set
\begin{equation}
C_\pm=\cd\ag+\cx\ax\pm\ar
\qquad {\rm and} \qquad
S=\sd\ag-\sx\ax\,,
\label{defhdiag}
\end{equation}
with the right hand sides containing the elements of $V_i$ in
eq.~(\ref{DiracMatrix}). They are specified in terms of the string profile
functions in eq.~(\ref{alphas}) of the following appendix.
The interaction Hamiltonian matrix elements read
\begin{eqnarray}
\langle 1r| H_{\rm int}|1s\rangle&=&
\ah\left(f_1^{(r)}f_1^{(s)}-g_1^{(r)}g_1^{(s)}\right)
-S\left(f_1^{(r)}g_1^{(s)}+g_1^{(r)}f_1^{(s)}\right)\cr
\langle 2r| H_{\rm int}|1s\rangle&=&
-C_{+}f_2^{(r)}g_1^{(s)}-C_{-}g_2^{(r)}f_1^{(s)}\cr
\langle 3r| H_{\rm int}|1s\rangle&=&
S\left(f_3^{(r)}f_1^{(s)}+g_3^{(r)}g_1^{(s)}\right)\cr
\langle 4r| H_{\rm int}|1s\rangle&=&
C_{-}f_4^{(r)}f_1^{(s)}+C_{+}g_4^{(r)}g_1^{(s)}
-\alp\left(f_4^{(r)}g_1^{(s)}-g_4^{(r)}f_1^{(s)}\right)\cr\cr
\langle 1r| H_{\rm int}|2s\rangle&=&
-C_{-}f_1^{(r)}g_2^{(s)}-C_{+}g_1^{(r)}f_2^{(s)}\cr
\langle 2r| H_{\rm int}|2s\rangle&=&
\ah\left(f_2^{(r)}f_2^{(s)}-g_2^{(r)}g_2^{(s)}\right)
+S\left(f_2^{(r)}g_2^{(s)}+g_2^{(r)}f_2^{(s)}\right)\cr
\langle 3r| H_{\rm int}|2s\rangle&=&
C_{+}f_3^{(r)}f_2^{(s)}+C_{-}g_3^{(r)}g_2^{(s)}
+\alp\left(f_3^{(r)}g_2^{(s)}-g_3^{(r)}f_2^{(s)}\right)\cr
\langle 4r| H_{\rm int}|2s\rangle&=&
-S\left(f_4^{(r)}f_2^{(s)}+g_4^{(r)}g_2^{(s)}\right)\cr\cr
\langle 1r| H_{\rm int}|3s\rangle&=&
S\left(f_1^{(r)}f_3^{(s)}+g_1^{(r)}g_3^{(s)}\right)\cr
\langle 2r| H_{\rm int}|3s\rangle&=&
C_{+}f_2^{(r)}f_3^{(s)}+C_{-}g_2^{(r)}g_3^{(s)}
-\alp\left(f_2^{(r)}g_3^{(s)}-g_2^{(r)}f_3^{(s)}\right)\cr
\langle 3r| H_{\rm int}|3s\rangle&=&
\ah\left(f_3^{(r)}f_3^{(s)}-g_3^{(r)}g_3^{(s)}\right)
-S\left(f_3^{(r)}g_3^{(s)}+g_3^{(r)}f_3^{(s)}\right)\cr
\langle 4r| H_{\rm int}|3s\rangle&=&
-C_{-}f_4^{(r)}g_3^{(s)}-C_{+}g_4^{(r)}f_3^{(s)}\cr\cr
\langle 1r| H_{\rm int}|4s\rangle&=&
C_{-}f_1^{(r)}f_4^{(s)}+C_{+}g_1^{(r)}g_4^{(s)}
+\alp\left(f_1^{(r)}g_4^{(s)}-g_1^{(r)}f_4^{(s)}\right)\cr
\langle 2r| H_{\rm int}|4s\rangle&=&
-S\left(f_2^{(r)}f_4^{(s)}+g_2^{(r)}g_4^{(s)}\right)\cr
\langle 3r| H_{\rm int}|4s\rangle&=&
-C_{+}f_3^{(r)}g_4^{(s)}-C_{-}g_3^{(r)}f_4^{(s)}\cr
\langle 4r| H_{\rm int}|4s\rangle&=&
\ah\left(f_4^{(r)}f_4^{(s)}-g_4^{(r)}g_4^{(s)}\right)
+S\left(f_4^{(r)}g_4^{(s)}+g_4^{(r)}f_4^{(s)}\right)\,.
\label{hintmat}
\end{eqnarray}
To keep the presentation simple we have omitted the radial integrals
on the right hand sides, {\it i.e.} they are understood
to be integrated with $\int_0^R\rho\,d\rho\,\left(\ldots\right)$.
In total this defines $8r_{\rm max}\times8r_{\rm max}$ matrix elements
of the interaction Hamiltonian. To populate the full Hamiltonian matrix,
we set
\begin{equation}
H^{(I)}_{r+2r_{\rm max},s}=\langle 2r |H_{\rm int}|1s\rangle
\label{offdiag}
\end{equation}
for $r,s=1,\ldots,2r_{\rm max}$, and $\epsilon_{r+2qr_{\rm max}}=\epsilon_r$
for $q=1,2,3$. This yields the
$8r_{\rm max}\times8r_{\rm max}$ matrix
\begin{equation}
H_{r,s}=\epsilon_r\delta_{rs}+H^{(I)}_{r,s} \qquad
r,s=1,\ldots,8r_{\rm max}\,,
\label{Hmat}
\end{equation}
which is  diagonalized numerically by means of a Jacobi routine.

Once the radius $R$ and the momentum cutoff $\Lambda$ are large enough
the \emph{true} bound state spectrum should become stable against further increase of
these parameters. Typical values are $\Lambda\approx 8m$ and $R=75/m$, 
so that our free basis comprises about 400 energy eigenvalues,
each with fourfold degeneracy and the
Hamiltonian matrix has $1600\times1600$ entries for the lowest angular
momentum. For wider string profiles, bound states occur in higher and
higher angular momentum channels. For instance, bound states appear
for up to $\ell = 5$ when $w_H\approx 6/m$, while narrow widths
$w_H\le1/m$ only induce bound states in the channel $\ell=-n=-1$, {\it
i.e.} the effective $S$--wave channel.

We have verified the gauge independence of the bound state energies by 
checking that they are insensitive to variations in the shape of the 
gauge transformation profile $\xi(\rho)$. Also the zero mode in the
$\ell=-n=-1$ channel is observed for $\xi_1=\sfrac{\pi}{2}$ regardless
of the values of the width parameters.

\section{Scattering problem}
\label{appB}

In this appendix we describe the scattering solutions to the Dirac 
equation~(\ref{DiracMatrix}).  To this end we write
the Dirac Hamiltonian, eq.~(\ref{eqDirac}) in terms of $4 \times 4$ matrices
and derive the differential equation for the Jost function.

\subsection{Differential equation for Jost function}
\label{appB1}

The derivative operators as well as the angular barriers are contained in the
diagonal $4 \times 4$ matrices
\begin{eqnarray}
D_u &\equiv& \partial_\rho \,\ID - \frac{1}{\rho}\,\mathrm{diag}
\big(-(\ell+n+1)\,,\,-(\ell+1)\,,\,\ell+n\,,\,\ell\big) \nonumber \\[2mm]
D_d &\equiv& \partial_\rho \,\ID- \frac{1}{\rho}\,\mathrm{diag}
\big(\ell+n\,,\,\ell\,,\, -(\ell+n+1)\,,\,-(\ell+1)\big)
\end{eqnarray}
where $C = \mathrm{diag}(-1,-1,1,1)$. For the interactions, we must compute
the matrix elements of the various pieces in eq.~(\ref{eqDirac}) within the
grand spin basis eq.~(\ref{eq:GSstates}). The explicit expressions for
the emerging radial functions are listed in eqs.~(\ref{eq:GSspinors})
and~(\ref{vecnot}), {\it cf.} eq.~(\ref{eq:GSspinors_app}). After some lengthy algebra,
the interaction matrices in the system (\ref{DiracMatrix}) can written in
terms of simpler sub--matrices,
\begin{equation}
\begin{array}{ll}
V_{uu}=\begin{pmatrix}
H & G_{+} \cr
G_{-} & H \end{pmatrix} & \qquad
V_{dd}=\begin{pmatrix}
-H & G_{-} \cr
G_{+} & -H \end{pmatrix} \cr\cr
V_{ud}=-\begin{pmatrix}
G_{+} & P \cr
P & G_{-} \end{pmatrix} & \qquad
V_{du}=-\begin{pmatrix}
G_{-} & -P \cr
-P & G_{+} \end{pmatrix}\,,
\end{array}
\label{compactmatrices}
\end{equation}
where the $2\times2$ submatrices are
\begin{eqnarray}
H &=& \alpha_H\,\begin{pmatrix} 1 & 0 \cr 0 & 1 \end{pmatrix}\,, \qquad\qquad
P = \alpha_p\, \begin{pmatrix} 0 & -1 \cr 1 & 0 \end{pmatrix}\,,
\nonumber \\[3mm]
G_{\pm} &=& \alpha_G\, \begin{pmatrix} \sin\Delta & \cos\Delta \cr \cos\Delta & -\sin\Delta
\end{pmatrix}
+ \alpha_\xi\,\begin{pmatrix} -\sin\xi & \cos\xi \cr \cos\xi & \sin\xi \end{pmatrix} \pm
 \alpha_r\,\begin{pmatrix} 0 & -1 \cr 1 & 0 \end{pmatrix}\,.
\label{defHGpmP}
\end{eqnarray}
The coefficients $\alpha_H$, $\alpha_p$, $\alpha_G$,
$\alpha_\xi$ and $\alpha_r$ are radial functions determined by the background
profiles $f_G, f_H$ and the gauge function $\xi$,
\begin{eqnarray}
\alpha_r(\rho) &=& \frac{1}{2}\,\frac{\partial \xi(\rho)}{\partial\rho} \nonumber \\[2mm]
\alpha_G(\rho) &=& \frac{n}{2\rho}\,f_G(\rho)\,\sin \Delta(\rho) \nonumber \\[2mm]
\alpha_\xi(\rho) &=& \frac{n}{2\rho}\,\big(f_G(\rho) - 1\big)\,\sin\xi(\rho) \nonumber \\[2mm]
\alpha_H(\rho) &=& m \,\big(f_H(\rho)\,\cos\Delta(\rho) - 1\big) \nonumber \\[2mm]
\alpha_P(\rho) &=& m f_H(\rho)\,\sin\Delta(\rho)\,.
\label{alphas}
\end{eqnarray}
Note that the new gauge function also enters via
$\Delta(\rho)  \equiv \xi_1 - \xi(\rho)$.
We now write the Dirac equation for the matrix fields defined in
eq.~(\ref{Smat1}) as
\begin{eqnarray}
\partial_\rho \mathcal{F} &=&
\left[\overline{\mathcal{M}}_{ff}+O_d\right]\cdot\mathcal{F}
+\mathcal{F}\cdot\mathcal{M}_{ff}^{(r)}
+\left[\overline{\mathcal{M}}_{fg}+kC\right]\cdot\mathcal{G}\cdot Z_d\\[2mm]
\partial_\rho \mathcal{G} &=&
\left[\overline{\mathcal{M}}_{gg}+O_u\right]\cdot\mathcal{G}
+\mathcal{G}\cdot\mathcal{M}_{gg}^{(r)}
+\left[\overline{\mathcal{M}}_{gf}-kC\right]\cdot\mathcal{F}\cdot Z_u\,,
\label{deqforGF}
\end{eqnarray}
where the $4 \times 4$ matrices without an overline
are purely kinematic,
\begin{eqnarray}
Z_u&=&{\rm diag}\, \left(
\frac{H_{\ell+n}(k\rho)}{H_{\ell+n+1}(k\rho)}\,,
\frac{H_{\ell}(k\rho)}{H_{\ell+1}(k\rho)}\,,
\frac{H_{\ell+n+1}(k\rho)}{H_{\ell+n}(k\rho)}\,,
\frac{H_{\ell+1}(k\rho)}{H_{\ell}(k\rho)}\right)
\nonumber \\[4mm]
Z_d&=&{\rm diag}\, \left(
\frac{H_{\ell+n+1}(k\rho)}{H_{\ell+n}(k\rho)}\,,
\frac{H_{\ell+1}(k\rho)}{H_{\ell}(k\rho)}\,,
\frac{H_{\ell+n}(k\rho)}{H_{\ell+n+1}(k\rho)}\,,
\frac{H_{\ell}(k\rho)}{H_{\ell+1}(k\rho)}\right)=\left(Z_u\right)^{-1}
\nonumber \\[4mm]
O_u&=&\frac{1}{\rho}\,{\rm diag}\,
\left(-(\ell+n+1),-(\ell+1),\ell+n,\ell\right) \nonumber \\[4mm]
O_d&=&\frac{1}{\rho}\,{\rm diag}\,
\left(\ell+n,\ell,-(\ell+n+1),-(\ell+1)\right) \nonumber \\[4mm]
C &=& \mathrm{diag}(-1,-1,1,1)\,.
\label{defDZud}
\end{eqnarray}
The matrices multiplying $\mathcal{F}$ and $\mathcal{G}$ from the right
are also independent of the background potential,
\begin{equation}
\mathcal{M}_{ff}^{(r)}=-kC\cdot Z_d-O_d
\qquad {\rm and} \qquad
\mathcal{M}_{gg}^{(r)}=kC\cdot Z_u-O_u\,.
\label{kinmatrix}
\end{equation}
Genuine interactions from the string background are solely contained in
the overlined matrices in eq.~(\ref{deqforGF}). Using the same {$2\times2$}
matrix notation as above, we have explicitly
\begin{equation}
\begin{array}{ll}
\overline{\mathcal{M}}_{gg}=CV_{ud}
=\begin{pmatrix}
G_{+} & P \cr -P & -G_{-}
\end{pmatrix}  & \qquad
\overline{\mathcal{M}}_{ff}=-CV_{du}
=\begin{pmatrix}
-G_{-} & P \cr  -P & G_{+}
\end{pmatrix} \cr \cr
\overline{\mathcal{M}}_{gf}=\frac{1}{\kappa}CV_{uu}
=\frac{1}{\kappa}\begin{pmatrix}
-H & -G_{+} \cr  G_{-} & H
\end{pmatrix}& \qquad
\overline{\mathcal{M}}_{fg}=-\kappa C V_{dd}
=\kappa\begin{pmatrix}
-H & G_{-} \cr  -G_{+} & H
\end{pmatrix} \,.
\end{array}
\label{intmatrix}
\end{equation}
The solutions to the differential equations~(\ref{deqforGF}) subject to
the boundary conditions $\mathcal{F} \to \ID$ and
$\mathcal{G} \to \ID$ at $\rho \to \infty$ define the
scattering solution, eq.~(\ref{scatsol}), from which we extract the
scattering matrix as described in eq.~(\ref{Smat3}).

\subsection{Born series}

To set up the Born series defined in eq.~(\ref{defBorn}), we simply expand the
system of differential equations from the last section in powers of the
background potential, which only enters the overlined matrices.
At first order, we obtain
\begin{eqnarray}
\partial_\rho \mathcal{F}^{(1)} &=&
O_d\cdot \mathcal{F}^{(1)}+\mathcal{F}^{(1)}\cdot \mathcal{M}_{ff}^{(r)}
+kC\cdot \mathcal{G}^{(1)}\cdot Z_d+\overline{\mathcal{M}}_{ff}
+\overline{\mathcal{M}}_{fg}\cdot Z_d\\[2mm]
\partial_\rho \mathcal{G}^{(1)} &=&
O_u\cdot \mathcal{G}^{(1)}+\mathcal{G}^{(1)}\cdot \mathcal{M}_{gg}^{(r)}
-kC\cdot \mathcal{F}^{(1)}\cdot Z_u+\overline{\mathcal{M}}_{gg}
+\overline{\mathcal{M}}_{gf}\cdot Z_u\,.
\label{firstborn}
\end{eqnarray}
The matrices $\mathcal{M}_{\ldots}^{(r)}$ do not contain the
interactions and are thus of order zero. In the same way we obtain the second
order equations,
\begin{eqnarray}
\partial_\rho \mathcal{F}^{(2)} &=&
O_d\cdot \mathcal{F}^{(2)}+\mathcal{F}^{(2)}\cdot \mathcal{M}_{ff}^{(r)}
+kC\cdot \mathcal{G}^{(2)}\cdot Z_d+\overline{\mathcal{M}}_{ff}\cdot \mathcal{F}^{(1)}
+\overline{\mathcal{M}}_{fg}\cdot \mathcal{G}^{(1)}\cdot Z_d\\[2mm]
\partial_\rho \mathcal{G}^{(2)} &=&
O_u\cdot \mathcal{G}^{(2)}+\mathcal{G}^{(2)}\cdot \mathcal{M}_{gg}^{(r)}
-kC\cdot \mathcal{F}^{(2)}\cdot Z_u+\overline{\mathcal{M}}_{gg}\cdot \mathcal{G}^{(1)}
+\overline{\mathcal{M}}_{gf}\cdot \mathcal{F}^{(1)}\cdot Z_u \,.
\label{secondborn}
\end{eqnarray}
With these Jost--like matrices, the Born series for the $S$--matrix
is $\mathcal{S}=\ID+\mathcal{S}^{(1)}+\mathcal{S}^{(2)}+\ldots\,$ with
\begin{eqnarray}
\mathcal{S}^{(1)}&=&\lim_{\rho\to0}\left\{ \mathcal{H}_u^{-1}\cdot\left[
\mathcal{F}^{(1)^*}-\mathcal{F}^{(1)}\right]\cdot\mathcal{H}_u^*\right\}\, 
\nonumber \\[2mm]
\mathcal{S}^{(2)}&=&\lim_{\rho\to0}\left\{ \mathcal{H}_u^{-1}\cdot\left[
\mathcal{F}^{(1)}\cdot\left(\mathcal{F}^{(1)}-\mathcal{F}^{(1)^*}\right)
+\mathcal{F}^{(2)^*}-\mathcal{F}^{(2)}\right]\cdot\mathcal{H}_u^*\right\}
\label{Smatborn}
\end{eqnarray}
and similarly for $\mathcal{G}_i$ with
$\mathcal{H}_u \to \mathcal{H}_d$.
The Born expanded eigenphase shifts are now simply given, to first and second
order, by
\begin{equation}
\delta_\ell^{(1)}=-\frac{1}{2}\,{\rm tr}\left[{\sf Im}
\left(\mathcal{S}^{(1)}_1\right)\right]
\qquad {\rm and} \qquad
\delta_\ell^{(2)}=-\frac{1}{4}\,{\rm tr}\left[{\sf Im}
\left(\mathcal{S}^{(1)}\cdot\mathcal{S}^{(1)}+2\mathcal{S}^{(2)}\right)\right]\,.
\label{phaseborn}
\end{equation}

The third and fourth order pieces will be treated as part of the
{\it fake boson} formalism discussed below in section \ref{app_fake}.

\subsection{Analytic continuation}

We describe the continuation to imaginary momenta for the case where
$\epsilon=\sqrt{k^2+m^2}$; the second Riemann sheet 
($\epsilon=-\sqrt{k^2+m^2}$) works analogously.
The analytic continuation concerns the Hankel functions, which turn
into modified Bessel functions, $Z_u\to Y_u$ and $Z_d\to Y_d$, with
\begin{equation}
Y_u={\rm diag}\, \left(
\frac{K_{\ell+n}(t\rho)}{K_{\ell+n+1}(t\rho)}\,,
\frac{K_{\ell}(t\rho)}{K_{\ell+1}(t\rho)}\,,
-\frac{K_{\ell+n+1}(t\rho)}{K_{\ell+n}(t\rho)}\,,
-\frac{K_{\ell+1}(t\rho)}{K_{\ell}(t\rho)}\right) =
-\left(Y_d\right)^{-1}\,.
\label{defZ12}
\end{equation}
Furthermore, the kinematic coefficient
turns into a pure phase
\begin{equation}
\kappa\to z_\kappa=\frac{m+i\sqrt{t^2-m^2}}{t} \,.
\label{kappaAC}
\end{equation}
The system of differential equations for imaginary momentum then becomes
\begin{eqnarray}
\partial_\rho \mathcal{F} &=&
\left[\overline{\mathcal{M}}_{ff}+O_d\right]\cdot\mathcal{F}
+\mathcal{F}\cdot\mathcal{M}_{ff}^{(r)}
+\left[\overline{\mathcal{M}}_{fg}-tC\right]\cdot\mathcal{G}\cdot Y_d\\[2mm]
\partial_\rho \mathcal{G} &=&
\left[\overline{\mathcal{M}}_{gg}+O_u\right]\cdot\mathcal{G}
+\mathcal{G}\cdot\mathcal{M}_{gg}^{(r)}
+\left[\overline{\mathcal{M}}_{gf}+tC\right]\cdot\mathcal{F}\cdot Y_u\,.
\label{deqforGFAC}
\end{eqnarray}
with the boundary conditions that $\mathcal{F}$ and $\mathcal{G}$ both
approach unity at $\rho \to \infty$. For simplicity, we have omitted the
momentum arguments in the radial wave--functions $\mathcal{F}$ and $\mathcal{G}$
and also used the same symbol as in the case of real momenta,
eqs.~(\ref{deqforGF}). The coefficient matrices in the
differential equations are slightly modified:
\begin{equation}
\begin{array}{ll}
\mathcal{M}_{gg}^{(r)}=-t\,C\cdot Y_u-O_u & \qquad
\mathcal{M}_{ff}^{(r)}=t\,C\cdot Y_d-O_d \\[2mm]
\overline{\mathcal{M}}_{gf}=z_\kappa
\begin{pmatrix}
-H & -G_{+} \cr  G_{-} & H
\end{pmatrix}  & \qquad
\overline{\mathcal{M}}_{fg}=
-z_\kappa^*
\begin{pmatrix}
-H &  G_{-} \cr -G_{+} & H
\end{pmatrix} \,,
\end{array}
\label{bornmatricesAC}
\end{equation}
while $\overline{\mathcal{M}}_{gg}$ and $\overline{\mathcal{M}}_{ff}$
are the same as on the real axis.

Unlike the Schr\"odinger problem, the differential equations in the present
case do not become real on the imaginary axis. Rather, charge conjugation
$\epsilon\to-\epsilon$ induces complex conjugation. It is therefore not 
surprising that the na{\"\i}ve extrapolation 
$\lim_{\rho\to0}\det(\mathcal{F})$
does not give a real result.  Instead, we find numerically that
$F=G^*$ with the imaginary part being independent
of angular momentum for a given value of $t$. The origin for this imaginary
part lies in the subtle definition of the Jost function via the Wronskian 
between the Jost solution, {\it i.e.} $\mathcal{F}$ or $\mathcal{G}$, and 
the regular solution, which satisfies momentum independent
boundary conditions at the origin. The momentum independence of
these boundary conditions ensures that the regular solution is an analytic
function of complex momentum. Analyticity of the Jost solution, on the
other hand, is guaranteed by the non--singular behavior of the
interaction potentials, which is in turn a consequence of the
boundary conditions on the profile function $\xi(\rho)$.
At the origin, the Higgs field differs from its vacuum expectation value
(it actually vanishes), which modifies the relative weight of
the upper and lower Dirac components. More precisely, the non--diagonal 
elements of the matrices in eq.~(\ref{bornmatricesAC}) vanish at the 
origin and the eight differential equations decouple with respect to the 
spin and weak isospin index on the radial functions in eq.~(\ref{vecnot}).
For real momenta $k$, a typical solution in the vicinity of
$\rho=0$ then looks like~\cite{Bordag:2003at}
\begin{equation}
\begin{pmatrix} f_4 \cr g_4 \end{pmatrix}
\sim \left(\frac{k}{q}\right)^{l}
\begin{pmatrix} \sqrt{E+mc_\Delta f_H(0)}\, J_l(q\rho) \cr\cr
\sqrt{E-m c_\Delta f_H(0)}\, J_{l+1}(q\rho) \end{pmatrix}
\label{regsol}
\end{equation}
with $q=\sqrt{E^2-(m c_\Delta f_H(0))^2}$
and similar dependencies for the other six radial functions.
The square--root coefficients cause the proper definition of the
logarithmic Jost function, $\nu(t)$, to be
\begin{equation}
{\rm exp}\left[\nu(t)\right]
=\left(\frac{\tau-im}{\tau-im c_\Delta f_H(0)}\right)^2 \lim_{\rho\to0}
{\rm det}(\mathcal{F})
=\left(\frac{\tau+im}{\tau+im c_\Delta f_H(0)}\right)^2 \lim_{\rho\to0}
{\rm det}(\mathcal{G})
\label{defJost}
\end{equation}
with $\tau=\sqrt{t^2-m^2}$. The power of two occurs
because we compute the determinant of a $4\times4$ matrix. Notice that
this redefinition not only cancels the imaginary parts, but
also modifies the real part. Furthermore it avoids the logarithmic singularity
in ${\rm ln}\left[\lim_{\rho\to0} {\rm det}(\mathcal{F})\right]$ otherwise observed
numerically at $t\sim m$. Since $f_H$ is part of the interaction, the
correction prefactor in eq.~(\ref{defJost}) also contributes to the Born
series. To make this explicit, we write
\begin{eqnarray}
{\rm ln}\left(\frac{\tau-im}{\tau-im c_\Delta f_H(0)}\right)&=&
{\rm ln}\left(\frac{\tau-im}{\tau-i(\alpha_H(0)+m)}\right)\\[2mm]
&=&\frac{i\alpha_H(0)}{\tau-im}
-\frac{1}{2}\left(\frac{\alpha_H(0)}{\tau-im}\right)^2+\ldots\,,
\label{expandfactor}
\end{eqnarray}
and subsequently set $\alpha_H(0)=-m$. The Born expansion of the
remaining determinant in eq.~(\ref{defJost}) is constructed as for real
momenta by iterating the differential equation~(\ref{deqforGFAC}) in the
interaction $\overline{\mathcal{M}}_i$.

Numerically, we integrate the differential equations~(\ref{deqforGF}),
(\ref{deqforGFAC}), their Born expansions and the fake boson
analog\footnote{The boundary
condition is $\overline{\nu}_\ell(t,\infty)=
\partial_\rho \overline{\nu}_\ell(t,\infty)=0$. The second order
contribution required in eq.~(\ref{finalJost}) is obtained from the
expansion $\overline{\nu}_\ell=\overline{\nu}_\ell^{(1)}+
\overline{\nu}_\ell^{(2)}+\ldots$, where the superscript labels the
order in $V(\rho)$.}
\begin{equation}
\partial_\rho^2 \overline{\nu}_\ell(t,\rho) =2 t L_\ell(t\rho)
\partial_\rho \overline{\nu}_\ell(t,\rho)
-\overline{\nu}_\ell^2(t,\rho)+V(\rho)
\quad \mbox{with}\quad
L_\ell(z)=\frac{K_{\ell+1}(z)}{K_\ell(z)}-\frac{\ell+\frac{1}{2}}{z}\,
\label{fakeanalog}
\end{equation}
\noindent
from some large radius $\rho_{\rm max}\sim 4\rho_0$ to $\rho_{\rm min}\sim0$
with the boundary condition $\mathcal{F}(\rho_{\rm max},k)=\ID$, and
identify $\lim_{\rho\to0}\mathcal{F}(\rho,k)
=\mathcal{F}(\rho_{\rm min},k)$. Alternatively, this identification
can also  be obtained from the derivative of the
wave--function. Furthermore, a differential equation is formulated
for ${\rm ln}\,{\rm det} \mathcal{F}(\rho,k)^{-1}
\mathcal{F}(\rho,k)^\ast$ to avoid $2\pi$
ambiguities in the computation of the phase shift, $\delta_\ell(k)$,
{\it cf.} eq.~(\ref{delta}).

The computations for real momenta have been performed mainly for use in the
consistency tests on the unitarity of the scattering matrix
and the spectral sum rules~\cite{Graham:2001iv}. There is one more merit
of considering real momenta: Channels that include Hankel
functions with zero index ($\ell=-2,-1,0$) are particularly cumbersome
because regular and irregular solutions are difficult to separate in such
cases, because they go as a constant and ${\rm ln}(\rho)$,
respectively, at $\rho_{\rm min} \ll 1$, {\it cf.} eq.~(\ref{Smat3}).
As a consequence, $\rho_{\rm min}$ must be taken tiny in the problematic
channels to obtain the correct scattering matrix in eq.~(\ref{Smat3}).
On the real axis, the result can be checked against extracting the
$\mathcal{S}$--matrix from the derivative of the scattering
wave--function because $Y_0^\prime(\rho)\sim Y_1(\rho)$ diverges like
a power. For calculations on the imaginary axis, we
assume $\rho_{\rm min}\sim 10^{-60}$ and successively carry out
an extrapolation
\begin{equation}
\nu(\rho_{\rm min})=\nu_0 +\frac{a_1}{{\rm ln}(\rho_{\rm min})}
+\frac{a_2}{{\rm ln}^2(\rho_{\rm min})}\ldots\,\,,
\label{nuextra}
\end{equation}
for the Jost function in these channels. We test the final result,
{\it i.e.}~$\nu_0$, for stability against further changes of
$\rho_{\rm min}$ and also check the condition ${\sf Im}(\nu_0)=0$.
In the non--problematic channels $\ell \notin \{ -2, -1, 0 \}$,
it is sufficient to set $\rho_{\rm min}\sim 10^{-12}$ in order to represent
the origin.

We have also successfully tested our numerical results of the
scattering data against the reflection symmetry $\ell \to  -(\ell+2n)$.

\section{Feynman diagrams}
\label{appC}

In this appendix we describe the details of the computation of the
Feynman diagrams. We start from the series in
equation~(\ref{Feynmanseries}).  This computation involves three parts:
\begin{enumerate}
\item The contribution linear in $H_I$. This term
vanishes identically with the no--tadpole condition.
\item The piece quadratic in $H_I$. This term is quadratically
divergent at high momenta and must be carefully regularized to handle
the leading and subleading logarithmic divergences.
\item The contribution from terms cubic and quartic in $H_I$.
They are only logarithmicly divergent which makes the separation of
the finite parts simpler. Since the corresponding
Feynman diagrams are complicated to evaluate we employ the {\it fake boson}
methods to compute this part of the vacuum polarization energy.
\end{enumerate}

\noindent
We first consider the $\overline{\rm MS}$ scheme in which only the
bare divergences proportional
to
$$
-i\left(\frac{\mu}{m}\right)^{4-D}
\int \frac{d^Dl}{(2\pi)^D}\left(l^2-1+i\epsilon\right)^{-2}\,,
$$
are subtracted, and then determine the finite counterterm coefficients suitable
to implement the on--shell renormalization scheme.

\subsection{Second order contribution}

After imposing the no--tadpole condition\footnote{The $c_3$ type counterterm in
eq.~(\ref{Lcounter}) contains a term quadratic in the fluctuations about the
Higgs $vev$. Its finite contribution is essential to keep the pseudo--scalar
part of the Higgs field massless, {\it i.e.} the expansion of the coefficient
of $p(k)p(-k)$ starts at $\mathcal{O}(k^2)$.} we find the contribution to the
action functional up to second order in $H_I$ within the $\overline{\rm MS}$
scheme as
\begin{eqnarray}
\Delta \mathcal{A}&=&-\frac{1}{8\pi^2} \int\frac{d^4k}{(2\pi)^4}
\int_0^1 dx \, {\rm ln}\left[1-x(1-x)\frac{k^2}{m^2}\right] \cr
&& \hspace{0.2cm}\times {\rm tr}_I
\Big\{\hspace{-0.2cm}\left[m^2-x(1-x)k^2\right]
\left[\frac{1}{2} L(k)\cdot L(-k)
-3\left(h(k)h(-k)+p(k)p(-k)\right)\right] \cr
&&\hspace{1.5cm} +x(1-x)\left[k\cdot L(k)\, k\cdot L(-k)
-\frac{1}{2}k^2 L(k)\cdot L(-k)\right] \cr
&&\hspace{1.5cm}+2m^2 p(k)p(-k)
+im k\cdot L(k) p(-k) \Big\} \cr
&&+\frac{1}{8\pi^2} \int\frac{d^4k}{(2\pi)^4} \,
\frac{k^2}{6}\,{\rm tr}_I\left[h(k)h(-k)+p(k)p(-k)\right]\,.
\label{Aren2}
\end{eqnarray}
Here the fields with momentum arguments are the Fourier transforms of the
corresponding spatial fields in eq.~(\ref{hint2}).
Specifically, we introduce the notation
$k:=k^\mu=(k_0,k_\perp \hat{\vek{k}}_\perp+k_3\hat{\vek{z}})^\mu$
and $L:=L^\mu=(L_0,\vek{L}_\perp+L_3\hat{\vek{z}})^\mu$ with $L_0=L_3=0$.
As a result, we have
\begin{eqnarray}
h(k)&=&h(-k)=-(2\pi)^3\delta(k_0)\delta(k_3)h_0(k_\perp) \\[2mm]
p(k)&=&p^\dagger(-k)=-(2\pi)^3\delta(k_0)\delta(k_3)(-i)^n\,p_n(k_\perp)\,
I_P(\varphi_k)\cr\cr
\vek{L}(k)&=&-(2\pi)^3\delta(k_0)\delta(k_3)\sum_{i=1}^3
\left[l^{(i)}_\perp(k_\perp,\varphi_k) \hat{\vek{k}}
+l^{(i)}_\varphi(k_\perp,\varphi_k)
\widehat{\mbox{\boldmath$\varphi$}_k}\right]\,,
\label{ft0}
\end{eqnarray}
where $\varphi_k$ is the azimuthal angle in
momentum space. The coefficients are isospin matrices,
$$
\begin{array}{ll}
\hspace{1cm}
l^{(1)}_\perp= (-i)^{n-1}\alpha_r^{(-)}(k_\perp)I_P(\varphi_k) \hspace{2cm}
&l^{(1)}_\varphi= (-i)^{n}\alpha_r^{(+)}(k_\perp)I_P(\varphi_k)
\begin{pmatrix}-1 & 0 \cr 0 &1 \end{pmatrix}\cr
\hspace{1cm}
l^{(2)}_\perp=0
&l^{(2)}_\varphi=i\alpha_s(k_\perp)\begin{pmatrix}-1 & 0 \cr 0 &1 \end{pmatrix} \cr
\hspace{1cm}
l^{(3)}_\perp= (-i)^{n-1}\alpha_c^{(+)}(k_\perp)I_P(\varphi_k)
&l^{(3)}_\varphi=(-i)^{n}\alpha_c^{(-)}(k_\perp)I_P(\varphi_k)
\begin{pmatrix}-1 & 0 \cr 0 &1 \end{pmatrix}\,.
\hspace{1.0cm} \mbox{\refstepcounter{equation}\label{ft00}(\theequation)}
\end{array}
$$
The matrix $I_P$ is defined in eq.~(\ref{Ip}), here to be taken as a
function of the azimuthal angle in momentum space. The functions 
$h_0$, $p_n$, $\alpha_r^{(-)}, \ldots$ are the Fourier transforms
\begin{eqnarray}
h_0(k)&=&\int_0^\infty \rho d\rho\, \alpha_H(\rho)J_0(k\rho)
\nonumber \\[2mm]
p_n(k)&=&\int_0^\infty \rho d\rho\, \alpha_P(\rho)J_n(k\rho)
\nonumber \\[2mm]
\alpha_r^{(\pm)}(k)&=& \int_0^\infty \rho d\rho\, \alpha_r(\rho)
\left[J_{n+1}(k\rho)\pm J_{n-1}(k\rho)\right]
\nonumber\\[2mm]
\alpha_s(k)&=&2\int_0^\infty \rho d\rho\,
\left[\ag(\rho)\sd(\rho)-\ax(\rho)\sx(\rho)\right]J_1(k\rho)
\nonumber \\[2mm]
\alpha_c^{(\pm)}(k)&=& \int_0^\infty \rho d\rho\,
\left[\ag(\rho)\cd(\rho)+\ax(\rho)\cx(\rho)\right]
\left[J_{n+1}(k\rho)\pm J_{n-1}(k\rho)\right]\,.
\label{ft1}
\end{eqnarray}
Some of these terms can be conveniently combined,
\begin{eqnarray}
\sum_{i=1}^3l^{(i)}_\perp &=& (-i)^{n-1}
\left[\alpha_r^{(-)}+\alpha_c^{(+)}\right]I_P(\varphi_k)\\[3mm]
\sum_{i=1}^3l^{(i)}_\varphi&=&i\alpha_s\begin{pmatrix}-1 & 0 \cr 0 &1 \end{pmatrix}
+(-i)^{n}\left[\alpha_r^{(+)}+\alpha_c^{(-)}\right]I_P(\varphi_k)
\begin{pmatrix}-1 & 0 \cr 0 &1 \end{pmatrix}\,.
\label{collectft}
\end{eqnarray}
Then we find the second order contribution to the energy
\begin{eqnarray}
\Delta E^{(2)}_{\rm FD}&=&\int_0^\infty \frac{k dk}{4\pi} \,\Bigg\{
\frac{k^2}{3}\left(h_0^2+p_n^2\right)
+4m^2I_1p_n^2+2mkI_1\left(\alpha_c^{(+)}+\alpha_r^{(-)}\right)p_n \cr
&&\hspace{0.7cm}
+k^2I_2\left[\left(\alpha_c^{(+)}+\alpha_r^{(-)}\right)^2
\hspace{-0.1cm}-\left(\alpha_c^{(-)}+\alpha_r^{(+)}\right)^2
\hspace{-0.1cm}-\left(\alpha_s\right)^2\right]
\cr
&&\hspace{0.7cm}-\left(m^2I_1+k^2I_2\right)
\left[6h_0^2+6p_n^2+\left(\alpha_c^{(+)}+\alpha_r^{(-)}\right)^2
+\left(\alpha_c^{(-)}+\alpha_r^{(+)}\right)^2+\left(\alpha_s\right)^2\right]
\Bigg\}\,,\qquad
\label{fdint0}
\end{eqnarray}
with the Feynman--parameter integrals ($\eta=k/m$)
\begin{eqnarray}
I_1&=&\int_0^1 dx \, {\rm ln}\left[1+x(1-x)\eta^2\right]
=\frac{2}{\eta}\sqrt{4+\eta^2}\,
{\rm arsinh}\left(\sfrac{\eta}{2}\right)-2\,,\label{fdint1}
\nonumber \\[2mm]
I_2&=&\int_0^1 dx \, x(1-x)\,{\rm ln}\left[1+x(1-x)\eta^2\right]
=\frac{\sqrt{4+\eta^2}}{3\eta^3}\left[\eta^2-2\right]
{\rm arsinh}\left(\sfrac{\eta}{2}\right)+\frac{2}{3\eta^2}-\frac{5}{18}\,.
\end{eqnarray}

\subsection{Fake boson method}
\label{app_fake}

We have already discussed the spectral part of the
fake boson approach in eq.~(\ref{fakeanalog}). Here we focus on the
Feynman diagram part. First we need to determine the logarithmicly
divergent contribution to action from the third and fourth order
Feynman diagrams. They can be parameterized by a radial integral,
\begin{eqnarray}
c_F&=&\int_0^\infty \rho\,d\rho\,\Big\{
\left(\alpha_H^2+\alpha_P^2\right)
\left(\alpha_H^2+\alpha_P^2+4m\alpha_H\right)
+4\ar\left(\ah\alp^\prime-\alp\ah^\prime\right)
\nonumber \\[2mm]
&&\hspace{2cm}
+4\left(\ar^2+\ag^2+\ax^2+2\ax\ag\cy\right)
\left(\alpha_H^2+\alpha_P^2+2m\alpha_H\right)
\nonumber \\[2mm]
&&\hspace{2cm}
-\frac{64}{3}\ar^2\left(\ag^2+\ax^2+2\ax\ag\cy\right)
-\frac{8}{3}\frac{n^2}{\rho^2}\ar f_G^\prime \sy\sx\sd
\nonumber \\[2mm]
&&\hspace{2cm}
-4\frac{n}{\rho}\alp\left[\ah\left(\ag\cd+\ax\cx\right)
+\alp\left(\ag\sd-\ax\sx\right)\right]\Big\}\,,
\label{lfctcoef}
\end{eqnarray}
where primes denote derivatives with respect to the radial coordinate.
With this  radial integral the divergence reads, in dimensional regularization,
\begin{equation}
\mathcal{A}^{\rm (div)}_{3,4}= \pi  c_F\, TL\,
\left[i\left(\frac{\mu}{m}\right)^{4-D}
\int\frac{d^Dl}{(2\pi)^D}\left(l^2-1+i\epsilon\right)^{-2}\right]\,.
\label{logdiv}
\end{equation}
Here $T$ and $L$ are the (infinite) lengths of the time and
$z$--axis intervals, respectively.

\smallskip
A boson field that fluctuates about a background potential
$V(\rho)=m^2\frac{\rho}{\rho_0}\, {\rm e}^{-2\rho/\rho_0}$
causes a similar logarithmic divergence for its vacuum polarization
energy at \emph{quadratic} order. In fact, the only replacement in
eq.~(\ref{logdiv}) is $c_F \to c_B$ with
\begin{equation}
c_B=\frac{1}{4}\int_0^\infty \rho\, d\rho\, V^2(\rho)
=\frac{3m^4\rho_0^2}{512}\,.
\label{boslogdiv}
\end{equation}
As for the spectral part, eq.~(\ref{finalJost}), we
rescale the fake boson potential with the strength of the fermionic
divergence $c_F$ so that we are only left with the finite part of the second
order (boson) Feynman diagram,
\begin{equation}
\Delta E_{\rm B}=-\frac{c_F}{c_B}
\int_0^\infty \frac{k dk}{16\pi} \, I_1 V_0^2\,.
\label{fdbos}
\end{equation}
In this equation, the Fourier transform of the fake boson background is
\begin{eqnarray}
V_0(k)&=&\int_0^\infty \rho\, d\rho\, V(\rho)J_0(k\rho)=
m^2\rho_0^2\,\frac{8-k^2\rho_0^2}
{\left[4+k^2\rho_0^2\right]^{\frac{5}{2}}}\,.
\label{ftbos}
\end{eqnarray}
For the numerical test mentioned after eq.~(\ref{VERYMASTER}) we
vary $\rho_0$ and verify that the vacuum polarization energy does not change.

\subsection{On--shell renormalization}
\label{app_onshell}

We parameterize the counterterm coefficients in dimensional regularization
\begin{equation}
c_s=-i\left(\frac{\mu}{m}\right)^{4-D}
\int \frac{d^Dl}{(2\pi)^D}\left(l^2-1+i\epsilon\right)^{-2}
+\overline{c}_s\,,
\label{finitecoef}
\end{equation}
for $s=1,\ldots,4$. In the $\overline{\rm MS}$
scheme, $\overline{c}_1 = \overline{c}_2 = \overline{c}_4 = 0$,
while the on--shell conditions discussed in the main text yield
\begin{eqnarray}
\overline{c}_1&=&
-\frac{1}{12}\frac{g^2}{\left(4\pi\right)^2}
\Bigg\{1+6\int_0^1 dx x\left(1-x\right)
\left[2{\rm ln}\left[1-x(1-x)\mu_W^2\right]
-\frac{x(1-x)\mu_W^2}{1-x(1-x)\mu_W^2}\right]\Bigg\}
\nonumber \\[4mm]
\overline{c}_2&=&
\frac{2f^2}{\left(4\pi\right)^2}
\left\{\frac{2}{3}+6\int_0^1 dx x\left(1-x\right)
{\rm ln}\left[1-x(1-x)\mu_H^2\right]\right\}
\nonumber \\[4mm]
\overline{c}_4&=&
-\frac{f^4}{2\left(4\pi\right)^2}
\left\{\mu_H^2+6\int_0^1 dx \,
{\rm ln}\left[1-x(1-x)\mu_H^2\right]\right\}\,.
\label{finitecounter}
\end{eqnarray}
We recall that the no--tadpole condition implies
$\overline{c}_3=m^4 / (4 \pi^2\,v^2) = f^2 m^2 / (4 \pi^2)$.
In this scheme, the pole position of the gauge boson field is not prescribed
but rather becomes a prediction. We find an implicit equation for the gauge
boson mass $\mu_W=m_W/m$:
\begin{eqnarray}
\mu_W^2&=&\frac{g^2}{2f^2}+\frac{g^2}{16\pi^2}\Bigg\{
\frac{2}{3}-\mu_W^2\left[\frac{1}{6}
-\mu_W^2\int_0^1dx\, \frac{x^2(1-x)^2}{1-x(1-x)\mu_W^2}\right]
\nonumber \\[2mm]
&&\hspace{2cm}
+6\int_0^1dx x(1-x){\rm ln}\left[1-x(1-x)\mu_W^2\right]
-\int_0^1dx\, {\rm ln}\left[1-x(1-x)\mu_W^2\right]\Bigg\}\,.\qquad
\label{implicitmass}
\end{eqnarray}

\section{Landau ghost estimate}

In the present treatment (without gauge boson loops) our model is not
asymptotically free. This results in unphysical poles of the renormalized 
propagators at large space--like momenta. These so-called \emph{Landau poles} 
are not real singularities but rather indicate the breakdown
of our treatment in certain momentum or parameter regimes. In the
present model the problem has a notable effect only for narrow
background profiles and/or large coupling constants. We have implemented a
procedure similar to that of ref.~\cite{Hartmann:1994ai} to verify
{\it a posteriori} that the interesting configurations do not suffer
from this unphysical effect.

Specifically, we write the renormalized quadratic contribution to the
energy per unit length coming from the pseudoscalar component of the Higgs
as
$$
\frac{v^2}{2}\int \frac{d^2q}{(2\pi)^2}\,
{\rm tr}\Big[p(q)\,p(-q)\Big] G_p^{-1}(q^2)
$$
which involves the corresponding (inverse) propagator for space--like momenta,
\begin{eqnarray}
G_p^{-1}(q^2)&=&q^2+\frac{f^2N_C}{8\pi^2}\Bigg\{q^2-6q^2\int_0^1 dx\, x(1-x)\,
{\rm ln}\,\frac{m^2+x(1-x)q^2}{m^2-x(1-x)m_H^2}
\nonumber \\[2mm]
&&\hspace{2cm}
-2m^2\int_0^1 dx \,x(1-x)\, {\rm ln}\left[1+x(1-x)\frac{q^2}{m^2}\right]\Bigg\}\,.
\label{gpiinv}
\end{eqnarray}
In the vicinity of the Landau pole ($q^2\sim m_G^2$) this propagator has
the expansion
\begin{equation}
G_p^{-1}(q^2)\sim\frac{1}{Z_G}\left(q^2-m_G^2\right)\,,
\label{ghost1}
\end{equation}
where
\begin{equation}
Z_G=\left(\frac{\partial G_p^{-1}(q^2)}{\partial q^2}\Bigg|_{q^2=m_G^2}
\right)^{-1}
\label{ghost2}
\end{equation}
is the residue of the pole. This allows us to remove the Landau pole explicitly
by introducing
\begin{equation}
\Delta_p^{-1}(q^2)=\left[\frac{1}{G_p^{-1}(q^2)}-\frac{Z_G}{q^2-m_G^2}\right]^{-1}\,.
\label{ghost3}
\end{equation}
We eliminate the artificial ghost contribution associated with the
Higgs field from the energy in chirally symmetric way
\begin{equation}
E^{(H)}_G=\int \frac{d^2q}{(2\pi)^2}\,
\left[\frac{1}{q^2}\Delta_p^{-1}(q^2)\right]\,
\left(D_\mu \phi\right)^{\rm T}(-q)\left(D^\mu \phi\right)(q)\,.
\label{ghost4}
\end{equation}
To study the effect of the Landau ghost, this quantity should be
compared to the
same contribution without the Landau ghost removal in eq.~(\ref{ghost3}), which
we call $E^{(H)}$.

In the same way, we can treat the gauge boson contribution to the renormalized
energy per unit length,
\begin{equation}
E^{(W)}=\frac{1}{2}\int \frac{d^2q}{(2\pi)^2}\,
{\rm tr}\,\Big[W_{\mu\nu}(q)W^{\mu\nu}(-q)\Big]\, G_W^{-1}(q^2)
\label{wghost1}
\end{equation}
where $W_{\mu\nu}(q)$ denotes the Fourier transform of the
field strength tensor for the static background, eqs.~(\ref{string_gauge})
and~(\ref{eqn:profile}), while
\begin{equation}
G_W^{-1}(q^2)=1+\frac{N_Cg^2}{16\pi^2}\int_0^1 dx \,
x(1-x)\,{\rm ln}\,\frac{m^2+x(1-x)q^2}{m^2-x(1-x)m_W^2}
\label{wghost2}
\end{equation}
describes the inverse gauge field propagator for space--like momenta. Again,
this propagator has a pole at $q^2=\bar{m}_G^2$ with residue $\bar{Z}_G$ which
we remove by defining the subtracted inverse propagator
\begin{equation}
\Delta_W^{-1}(q^2)=\left[\frac{1}{G_W^{-1}(q^2)}
-\frac{\bar{Z}_G}{q^2-\bar{m}_G^2}\frac{q^2}{\bar{m}_G^2}\right]^{-1}\,.
\label{wghost3}
\end{equation}
The Landau ghost eliminated gauge field energy then becomes
\begin{equation}
E^{(W)}_G=\frac{1}{2}\int \frac{d^2q}{(2\pi)^2}\,
{\rm tr}\Big[W_{\mu\nu}(q)W^{\mu\nu}(-q)\Big] \,\Delta_W^{-1}(q^2)\,.
\label{wghost4}
\end{equation}
Asymptotic freedom implies that the Landau poles at large spacelike momentum 
in the various propagators should disappear at any order in perturbation theory, 
and also in the full theory. We therefore expect that the
difference between $E^{(H)}+E^{(W)}$ and $E^{(H)}_G+E^{(W)}_G$ is small whenever
the effect of the unphysical Landau ghost in our model can be safely ignored.
For the model parameters that we found interesting, $g=0.72$ and
$f\approx2$, this condition is indeed satisfied since the
relative difference between $E^{(H)}+E^{(W)}$ and
$E^{(H)}_G+E^{(W)}_G$ is only a fraction of a percent even for narrow
configurations with $w_G = w_H = 0.1$.

To see that there are indeed background potentials where the Landau ghost
contribution is sizeable, we present the same comparison between
$E^{(H)}+E^{(W)}$ and $E^{(H)}_G+E^{(W)}_G$ for $g=f=10$ in table \ref{fig_landau}.
\begin{table}
\begin{tabular}{c|c|c}
$w_G=w_H$ & $E^{(H)}+E^{(W)}$ & $E^{(H)}_G+E^{(W)}_G$\cr
\hline
0.1& -15.597 & 3.220 \cr
0.5& -0.168 & 0.209 \cr
2.0& 0.041 & 0.082 \cr
4.0& 0.061 & 0.077 \cr
6.0& 0.068 & 0.077 \cr
8.0& 0.070 & 0.077 \cr
\end{tabular}
\caption{\label{fig_landau}\sf Landau ghost removal for $g=f=10$.
As an example we have chosen $\xi_1=0.3\pi$.}
\end{table}
We observe that the Landau ghost causes the well--known instability for narrow
configurations and large couplings~\cite{Ripka:1987ne}. However, for wider
configurations its effect is moderate even when the coupling is large.
It should be emphasized that the present approach to the Landau ghost problem
is only qualitative since the energy expressions (\ref{ghost1}) and
(\ref{wghost1}) are not rigorous. However, the present method
convinces us that the configurations discussed in the main
body of this article do not suffer from this problem.

\end{document}